\newif\ifarxiv
\newif\ifhassi
\arxivtrue
\hassitrue
\ifarxiv
\RequirePackage{snapshot}
\documentclass[
    reprint,
    superscriptaddress,
    amsmath,amssymb,
    aps,
    pra,
]{revtex4-2}
\setcitestyle{super}
\renewcommand{\doi}[1]{\href{http://doi.org/#1}{doi:#1}}
\else
\documentclass[journal=jacsat,suppinfo,bookmarksdepth=subparagraph]{achemso}
\newcommand{\doi}[1]{\href{http://doi.org/#1}{doi:#1}}
\fi
\newif\ifshowlabels
\showlabelsfalse
\newif\iffigdraft
\figdraftfalse
\newif\ifgentoc
\newif\ifeditmode
\editmodefalse
\gentocfalse
\newif\ifshowparagraphinline
\showparagraphinlinefalse
\ifeditmode\else
\showparagraphinlinefalse
\fi
\ifarxiv
\usepackage[sectionbib]{bibunits}
\defaultbibliographystyle{new}
\defaultbibliography{references}
\else
\fi
\usepackage{booktabs}
\usepackage{layouts}
\usepackage[acronyms]{glossaries}
\usepackage[procnames]{listings}
\usepackage{pygments} 
\usepackage{amsfonts}
\usepackage{blindtext}
\let\oldblindtext\blindtext
\renewcommand{\blindtext}{\textcolor{gray}{\oldblindtext}}
\usepackage{cmap}
\usepackage{datetime}
\usepackage{environ}
\usepackage{etoolbox}
\usepackage[utf8]{inputenc}
\DeclareUnicodeCharacter{03BD}{\ensuremath{\nu}}
\DeclareUnicodeCharacter{2212}{--}
\DeclareUnicodeCharacter{03BC}{\textmu}
\DeclareUnicodeCharacter{03B2}{\ensuremath{\beta}}
\DeclareUnicodeCharacter{03B1}{\ensuremath{\alpha}}
\DeclareUnicodeCharacter{00B0}{\ensuremath{^\circ}}
\DeclareUnicodeCharacter{00B1}{\ensuremath{\pm}}
\DeclareUnicodeCharacter{2192}{\ensuremath{\rightarrow}}
\DeclareUnicodeCharacter{03C3}{\ensuremath{\sigma}}
\DeclareUnicodeCharacter{03C4}{\ensuremath{\tau}}
\DeclareUnicodeCharacter{2009}{ }
\DeclareUnicodeCharacter{2080}{\ensuremath{_0}}
\DeclareUnicodeCharacter{2082}{\ensuremath{_2}}
\DeclareUnicodeCharacter{2084}{\ensuremath{_4}}
\DeclareUnicodeCharacter{2610}{\ensuremath{\Box}}
\DeclareUnicodeCharacter{1F449}{\ensuremath{\rightarrow}} 
\DeclareUnicodeCharacter{1F448}{\ensuremath{\leftarrow}} 
\iffigdraft
    \usepackage[draft]{graphicx}
\else
    \usepackage{graphicx}
\fi
\usepackage{hyphenat}
\usepackage{ifxetex}
\usepackage{mathrsfs}
\usepackage{calc}

\ifeditmode
\usepackage{pdfcomment}

\else
\newcommand{\pdfcomment}[2][]{}

\fi
\usepackage{hyperref}
\hypersetup{colorlinks=true, pdfstartview=FitV, linkcolor=linkcolor, citecolor=dbluecolor, urlcolor=dgreencolor, bookmarksdepth=subparagraph}
\renewcommand{\hypertarget}[2]{#2} 
\usepackage{bookmark}
\usepackage{sidecap}
\usepackage{soul}
\usepackage{textcomp}
\usepackage{url}
\usepackage{xcolor}
\usepackage{xspace}
\usepackage{multirow}
\usepackage{array}
\newcolumntype{R}{>{\vspace{2ex}\displaystyle}{r}}
\newcolumntype{L}{>{\displaystyle}{l}}
\definecolor{SUgrey}{HTML}{6F777D}
\definecolor{SUorange}{HTML}{D44500}
\definecolor{dbluecolor}{rgb}{.01,.02,0.29}
\definecolor{dgraycolor}{rgb}{0.50,0.50,0.50}
\definecolor{dgreencolor}{rgb}{0.0,0.4,0}
\definecolor{linkcolor}{cmyk}{0,0.7,0.5,0.5}
\usepackage{accsupp}    
\newcommand{\noncopynumber}[1]{%
    \BeginAccSupp{method=escape,ActualText={}}%
    #1%
    \EndAccSupp{}%
}

\usepackage[compact]{titlesec}
\ifshowparagraphinline
\titleformat{\paragraph}[runin]{\color{gray}\normalfont\bfseries\footnotesize}{}{3pt}{\hspace{0.75em}\ul{\footnotesize\thesubsection\theparagraph)\;}}[:]
\else
\renewcommand{\paragraph}[1]{\par\phantomsection\addcontentsline{toc}{paragraph}{#1}}
\fi
\makeatletter
\@ifundefined{linelabel}{%
\newcommand{\linelabel}[1]{}}{}
\makeatother
\makeglossaries
\renewcommand{\glossarysection}[2][]{}
\include{acronyms}
\renewcommand{\thesection}{\Roman{section}}
\renewcommand{\thesubsection}{\thesection.\arabic{subsection}}

\makeatletter
\renewcommand{\p@subsection}{}
\renewcommand{\p@subsubsection}{}
\makeatother
\newcounter{subfigure}[figure]
\newcounter{subfigurenonumber}
\newcounter{tempfigure}
\renewcommand\thesubfigure{
\arabic{tempfigure}\alph{subfigure}}
\renewcommand\thesubfigurenonumber{(\alph{subfigurenonumber})}
\newcommand{\subfig}[2]{%
    \setcounter{tempfigure}{\value{figure}}%
    \addtocounter{tempfigure}{1}%
    \refstepcounter{subfigure}%
    \setcounter{subfigurenonumber}{\value{subfigure}}%
    \expandafter\edef\csname ref#2\endcsname{\thesubfigurenonumber}
    \label{#1}%
    }
\newcommand{\titleblock}{
\newcommand\SUaffil{\affiliation{Department of Chemistry, Syracuse University, Syracuse, NY 13210, USA}}
\author{Romana Shathy}
\author{Alexandria Guinness}
\author{John M. Franck}
\SUaffil
\email{jmfranck@syr.edu}
\title{Exchange and Microviscosity of Dynamic Nanocompartments}
\date{Not for Final Submission Purposes: \today, \currenttime}
\date{\today}
}
\ifarxiv\else
    \titleblock
\fi
\newif\ifpoormancref
\poormancreffalse
\ifpoormancref
\usepackage[poorman]{cleveref}
\else
\usepackage{cleveref}
\fi
\crefname{equation}{Eq.}{Eqs.}
\crefname{table}{Table}{Tables}
\crefname{figure}{Fig.}{Figs.}
\crefname{section}{Sec.}{Sec.}
\crefname{subfigure}{Fig.}{Figs.}
\crefname{lstlisting}{listing}{listings}
\Crefname{lstlisting}{Listing}{Listings}

\usepackage{xr}
\ifshowlabels
\usepackage{refcheck}
\makeatletter
\newcommand{\refcheckize}[1]{%
  \expandafter\let\csname @@\string#1\endcsname#1%
  \expandafter\DeclareRobustCommand\csname relax\string#1\endcsname[1]{%
    \csname @@\string#1\endcsname{##1}\wrtusdrf{##1}}%
  \expandafter\let\expandafter#1\csname relax\string#1\endcsname
}
\def\@setmarginlbl{%
    \if@show@ref
        \if@labelled
            \set@fbox@par
            \if@unsdlbl
                \makebox[0pt][l]{\zero@height{$\,$\rotatebox{90}{\scalebox{0.6}{\mark@size
                {\bfseries\upshape?}\underline{\last@lbl}{k\bfseries\upshape?}}}}}%
            \else
                \makebox[0pt][l]{\zero@height{$\,$\rotatebox{90}{\scalebox{0.6}{\fbox{{\mark@size\last@lbl}}}}}}%
            \fi
        \else
            \if@show@unl@bld
                \makebox[0pt][l]{\zero@height{$\,$\rotatebox{90}{\scalebox{0.6}{\unl@bld@mark}}}}%
            \fi\fi
        \fi
        \global\@labelledfalse
    }
\def\@setnmmarginlbl{%
    \if@show@ref
        \set@fbox@par
        \if@unsdlbl
            \hbox to \textwidth{\makebox[0pt][r]{\rotatebox{90}{\scalebox{0.6}{\mark@size{\bfseries
                            \upshape?}$\langle$\last@lbl$\rangle${\bfseries
            \upshape?}}}$\,$}\hfill}%
        \else
            \hbox to \textwidth{\makebox[0pt][r]{\rotatebox{90}{\scalebox{0.6}{\mark@size$\langle$%
            \last@lbl$\rangle$}}$\,$}\hfill}%
        \fi
    \fi
    \global\@labelledfalse
}
\def\@bibitem@proceed@#1{%
    \@ifundefined{cit@#1}{\@warning@rc@{Unused bibitem `#1'}%
        \if@show@cite
            \gdef\@biblabel{\makebox[0pt][r]{\zero@height{\rotatebox{90}{\scalebox{0.6}{{\mark@size{\bfseries\upshape?}}%
                \underline{\@verbatim@{#1}}{\mark@size{\bfseries\upshape?}}}}$\,$}}%
            \@@@biblabel@@}%
        \fi
    }{%
        \if@show@cite
            \set@fbox@par
            \gdef\@biblabel{\makebox[0pt][r]{\zero@height{\rotatebox{90}{\scalebox{0.6}{\fbox{TESTTESTTEST\@verbatim@{#1}}}}$\,$}}\@@@biblabel@@}%
        \fi
}}%
\makeatother
\refcheckize{\cref}
\refcheckize{\Cref}
\fi
\begin{document}
\counterwithin{lstlisting}{section}
\newlength\myfigwidth
\setlength{\myfigwidth}{3.5in}
\ifarxiv
\titleblock
\fi

\begin{abstract}
\glspl{rm} provide a controlled system for studying
the unique structural and dynamic behaviors of confined water, thus
gaining a general insight into the behavior or water under confinement
and at interfaces. This study employs \gls{esr} to study the dynamics
inside \glspl{rm} and capitalizes on the fact that different
preparations of \glspl{rm} can precisely manipulate both the size of
water pools as well as whether or not different pools can come into
contact. A small spin label moiety, less than a half nanometer in size
tumbles in solution, offering insight into the rotational diffusion from
room temperature down to deeply super-cooled temperatures. Two different
spin probes can separately probe the dynamics of the interfacial layer
\emph{vs.} those at the core of the reverse micelle. The results provide
critical insights into how thermodynamic transitions affect the
stability and behavior of water in confined spaces, enhancing the
understanding of nucleation and ice crystal growth, as well as the
broader understanding of confined water's role in various chemical and
biological systems. The ability to more precisely understand and control
reaction dynamics and thermodynamics inside confined aqueous systems
across a broad range of temperatures will prove useful to fields ranging
from biophysics to the development of new nano-materials.

\end{abstract}

\glsresetall
\ifarxiv\begin{bibunit}\fi
\maketitle
\ifarxiv\glsresetall\fi
\section{Introduction}

Water, when confined, exhibits unique structural, dynamic, and
thermodynamic characteristics that differ markedly from its behavior in
bulk, making it an important subject for study
\cite{Cerveny2016ConWatMod,Beaton2023RapScrCor,Franck2019OveDynNuc,Agles2024StrDynWat,Monroe2020WatStrPro}.
Such confinement requires a macromolecular container that limits the
water to nanoscopic volumes. \glspl{rm}, \emph{i.e.} containers built of
surfactant dispersed in an apolar solvent, prove crucial for various
applications in chemistry,
biophysics,\cite{VanHorn2009RevMicEnc,Jolicoeur1974HydNitRad,Hauser1989IntWatSod,Avramiotis1999IntProLec,Malik2012MicMetNov,Cerveny2016ConWatMod}
and drug delivery \cite{Lin2023MemFusReva}. They not only confine
water within nanoscopic spaces but also possess the unique ability to
undergo fusion-fission processes
\cite{Guettari2017EffMicCol,Malik2012MicMetNov,Palmer2020NonMasExc,Rharbi2014FusFisInh},
along with tumbling and translational motion. Thus \glspl{rm} serve as
\emph{dynamic} nanocontainers. The interplay between the internal
molecular dynamics and the dynamics of the nanocontainers themselves
offers potential for manipulating the content and stability of guest
molecules or reactions contained within.

Electron Spin Resonance (\gls{esr}) offers detailed insights into the dynamics
of the local environments surrounding specially placed spin probes, and
this tool has been used to specifically study confined environments
\cite{Pan2018HowEnzOri}. For example, it has studied the hydration
layer of proteins in frozen solvent \cite{Li2022ConDepPro}. \gls{esr} has
also probed polymers, such as hydrogels or polymeric membranes
\cite{Goksel2019,Lawton2010,Pitt1993ESRSpePro,Windle1992ESRSpiPro},
and even complex nanostructures \cite{Kaser2022DomPhaTra}. \gls{esr} has
also been utilized to explore the dynamics of bulk water at deeply
supercooled temperatures \cite{Banerjee2009ESREviCoe}. Furthermore,
research has demonstrated that the rotational motion of a spin probe
becomes decoupled from the viscosity of bulk water at 277 K, with the
activation energy for the spin probe's rotation being higher than that
of water's viscosity below this temperature
\cite{Peric2013RotFouSmaa}.

In previous work, \gls{esr} investigated both the mobility of surfactants and
the dynamics of the aqueous environment within reverse micelles.
Previous literature \cite{Zuev2003EffProSol} employed the 4-SLBA
(4-(2-n-undecyl-3-oxyl-4,4,-dimethyloxazolidin-2-yl) butyric acid) spin
probe to study the mobility of the alkyl sidechains of the surfactant.
Literature going back a few decades agrees with the idea that the
mobility of a spin probe captured inside an \gls{rm} decreases as the size of
the \gls{rm} decreases \cite{Hauser1989IntWatSod}. Other early studies
showed that the mobility of the spin label changes as a function of the
water loading \cite{Haering1988ChaEleSpi}, and introduced a
core-shell model of bulk water surrounded by interfacial/bound water to
explain the correlation times observed \cite{Hauser1989IntWatSod}.
Some of this early work even employs spin labels to look at the changing
rotational correlation time of spin probes inside \glspl{rm} as $w_0$ changes
\cite{Hauser1989IntWatSod,Haering1988ChaEleSpi}; in other words,
what today would be understood as microviscosity. Tbl.~\ref{tbl:Haering}
further extends the analysis by applying a standard linewidth analysis
\cite{Kivelson1960TheESRLin,Avramiotis1999IntProLec}
(eq.~\ref{eq:tauFromLinewidth}) to the spectra presented in
\cite{Hauser1989IntWatSod,Haering1988ChaEleSpi}, in order to
determine the rotational diffusion as a function of water loading. This
analysis makes it clear that the rotational diffusion of the
surfactant-water interface spin label CAT-16 is slower than the aqueous
spin label DSTA. Furthermore, it illustrates that a decrease in the
water loading ($w_0$) further decreases the rotational correlation
time. Finally, by reporting on the local polarity, this previous work
demonstrates (see fig.~\ref{fig:HaeringCompositeFig}) how the hyperfine
splitting helps to determine the location of the spin label within a \gls{rm}
\cite{Haering1988ChaEleSpi}.

However, challenges and contradictions remain in understanding the
behavior of encapsulated water in \glspl{rm}. Early work on the \gls{esr} of \glspl{rm},
including studies using hydroxytempo, faced challenges due to the
equilibrium of spin labels between the water, surfactant layer, and
apolar dispersant \cite{Yoshioka1981TemDepMot}. Previous studies
\cite{Caldararu1998StrAspSel}, also indicated that the rotational
diffusion of spin labels follows Arrhenius temperature dependence at low
water loading (see fig.~\ref{fig:CaldararuTau}) \footnote{The findings
  presented in \cite{Caldararu1998StrAspSel} are attributed to
  aqueous spin labels; however, a comparison of the hyperfine couplings
  to those reported in \cite{Hauser1989IntWatSod} suggests that the
  spin probes might be residing within the interfacial solvent layer.}.
This observation seems to conflict with the well documented
nonlinearities in the viscosity of water as the temperature is lowered
from room temperature down to the dynamic crossover temperature near
222-225~K \cite{Cerveny2016ConWatMod}. A review of phase transitions
of the water inside \glspl{rm} reveals an interesting conflict. DSC studies
\cite{Boned1986ChaWatDis} report the freezing of water at
temperatures near the entry into what is now known as ``no-man's land''
near 230~K \cite{Cerveny2016ConWatMod} (specifically, DSC observed a
transition at a slightly higher temperature, -41°C). This freezing only
affects a sub-population of the water; in particular, freezing only
occurs for $w_0\ge 7$ and the enthalpy of fusion per mole of water
grows with $w_0$ \cite{Boned1986ChaWatDis}. To the authors'
knowledge, \gls{esr} had not previously reported on this transition inside
\glspl{rm}, despite the fact \gls{esr} has been used to explore the liquid water in
fissures of ice inside ``no man's land''
\cite{Banerjee2009ESREviCoe}. In contrast, \gls{nmr} studies
\cite{VanHorn2009RevMicEnc}, indicate the freezing of water at the
much higher temperature of 243 K. Earlier reports used \gls{esr} to observe a
similar effect \cite{Yoshioka1981TemDepMot}. In the current study,
\gls{esr} investigates both sample systems to offer insight into the reasons
for the discrepancy.

The present study employs specialized spin probes to retrieve unique
insights about the behavior of dynamic nanocontainers such as \glspl{rm}. Two
types of spin labels are employed: one (TEMPO-SO\textsubscript{4})
located in the middle of the \gls{rm} and another (CAT-16) in the interfacial
layer between the water and the surfactant. This study analyzes \gls{esr}
spectra to measure the microviscosity inside these nanocontainers over
wide temperature ranges. The present research investigated the mobility
of two spin labels using isooctane (melting point: -107.3°C $\approx$
166 K) and dodecane (melting point: -10°C $\approx 263\;\text{K}$) as
solvents. The study of \glspl{rm} in isooctane provides the information of
mobility of \gls{rm} and spin label over a wide temperature range, and
presents a new view on the process of cold-shedding of water. In
contrast, when the dodecane freezes, it immobilizes the \gls{rm}. This
provides the opportunity to distinguish between the mobility of the \gls{rm}
and spin label. As proves more important, it also suppresses fusion and
fission of the \gls{rm} aggregates.

The approach presented in the current article builds on these
foundations by using an anionic spin label that remains within the water
core, providing a clearer picture of the water dynamics within these
dynamic nanocontainers, and capitalizing on modern data-processing
tools, as well as spectroscopic simulation tools that offer more
detailed and reliable descriptions of the spin label motion. The
majority of this article utilizes a relatively qualitative analysis of
temperature-dependent studies of \glspl{rm} as a tool to bring together various
disparate observations from the literature and to paint a coherent
picture of how the water inside \glspl{rm} responds to changes in temperature.
Simulations are utilized to explore the behavior of the spin label when
the situation becomes more complicated, in particular where the water
inside the \glspl{rm} is supercooled.

\section{Experimental}

\subsection{\gls{esr}}

All EPR measurements were recorded at X-band (microwave frequency 9.4
GHz to 9.8 GHz) on a Bruker ELEXSYSII E500 spectrometer equipped with a
Bruker ER 4122 super-high Q (SHQE) resonator and SuperX bridge. A
variable-temperature accessory ER 4141VT enabled variable temperature
EPR measurements. In-house python scripts (based on pyspecdata
\cite{pyspecdata}) directly read, process, and plot the XEPR-format
spectral files, as well as export to ascii and extract the resonance
frequency for subsequent analysis in Altenbach's Multicomponent program;
similarly EasySpin can directly read the XEPR-format files. (Note that
both Multicomponent and EasySpin employ variants of the ``NLSL'' program
from the Freed lab \cite{budil1996nonlinear}.)

\subsection{Synthesis}

\subsubsection{\texorpdfstring{Synthesis ofTEMPO-SO\textsubscript{4}}{Synthesis of TEMPO-SO4}}

\begin{figure}
\centering
\includegraphics[width=\linewidth]{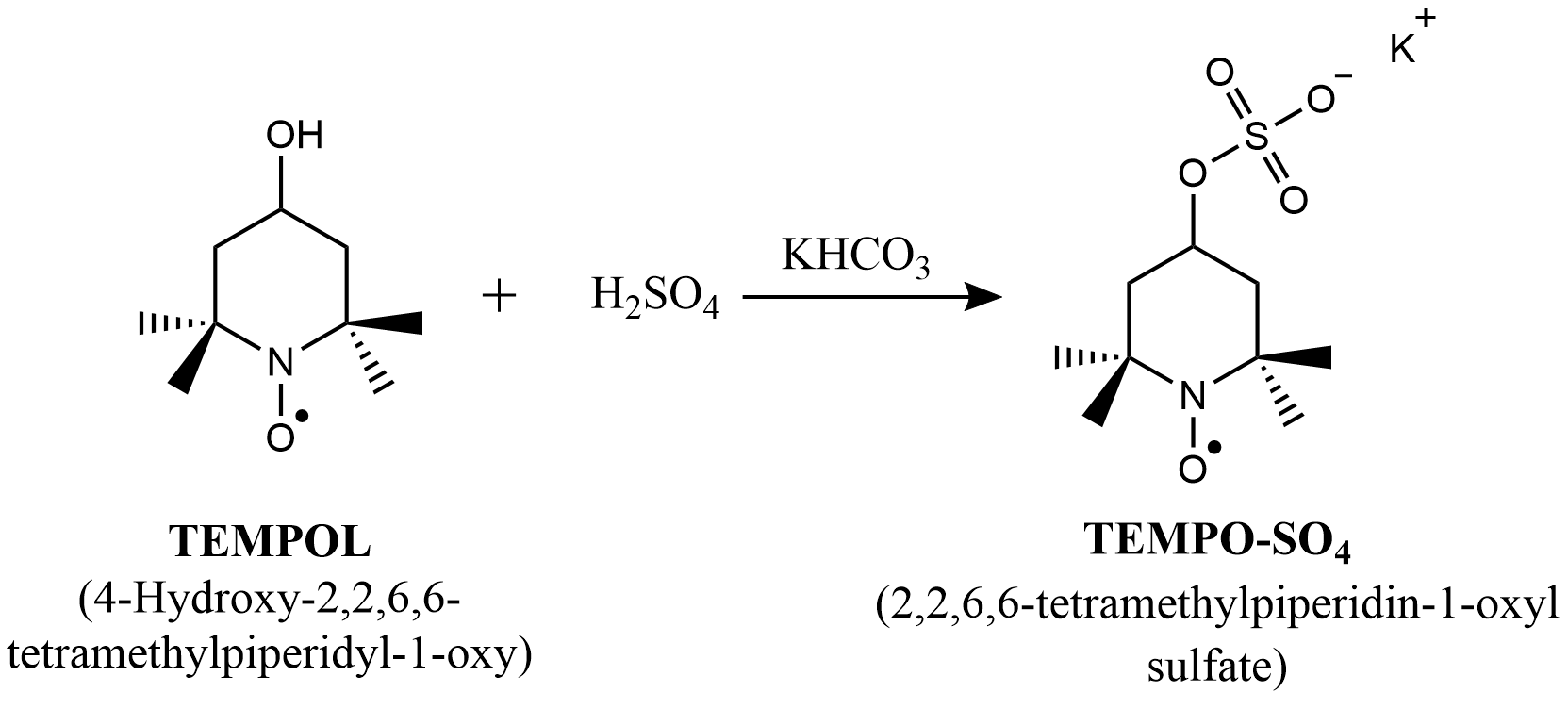}
\caption{Synthesis of
TEMPO-SO\textsubscript{4}}\label{fig:TEMPOSO4Reaction}
\end{figure}

The aqueous spin probe TEMPO-SO\textsubscript{4}
(2,2,6,6-tetramethylpiperidin-1-oxyl sulfate) was synthesized through
the sulfation of TEMPOL (4-Hydroxy-2,2,6,6-tetramethylpiperidyl-1-oxy)
following literature procedure \cite{Winsberg2017Aqu6TeCat}:

5 g (29.03 mmol) of TEMPOL ($\geq 98\%$, Toronto Reserch Chemicals)
was pestled, added slowly to 35 mL of concentrated sulfuric acid, and
stirred for 20 minutes at room temperature. The reaction mixture was
then added dropwise to 250 mL of 1.3 M KHCO\textsubscript{3} solution.
The aqueous phase was washed in an extraction funnel $5\times$ with 50
mL of ethyl acetate ($\geq 99.5\%$, Thermo Scientific). The water was
then removed with a rotary evaporator. The resulting yellow sticky solid
was added to 30mL acetone (HPLC grade, Fisher Scientific) and stirred
until only colorless solid remained. After vacuum filtration, most
acetone in the filtrate was removed \emph{via} rotary evaporation. The
resulting dark orange oil, after sitting for 72 hours in a desiccator
under reduced pressure, in the dark, became a dark red solid
(TEMPO-SO4). Some of this TEMPO-SO\textsubscript{4} was dissolved in
DMSO-d6 (99.9\%, Cambridge Isotope Laboratories) with a drop of phenyl
hydrazine (97\%, Acros Organics) to enable \gls{nmr}
(fig.~\ref{fig:NMR_TEMPOSO4}): \textsuperscript{1}H \gls{nmr}
(DMSO-d\textsubscript{6}, quenched with phenyl hydrazine, 400 MHz):
$\delta$ 1.04 (d, 12H), 1.31 (t, 2H), 1.92-1.97 (m, 2H), 4.32 (tt,
1H). A different aliquot was dissolved in water for \gls{esr} spectroscopy
that verified the presence of nitroxide.

\subsubsection{Synthesis of CAT-16}

\begin{figure}
\centering
\includegraphics[width=\linewidth]{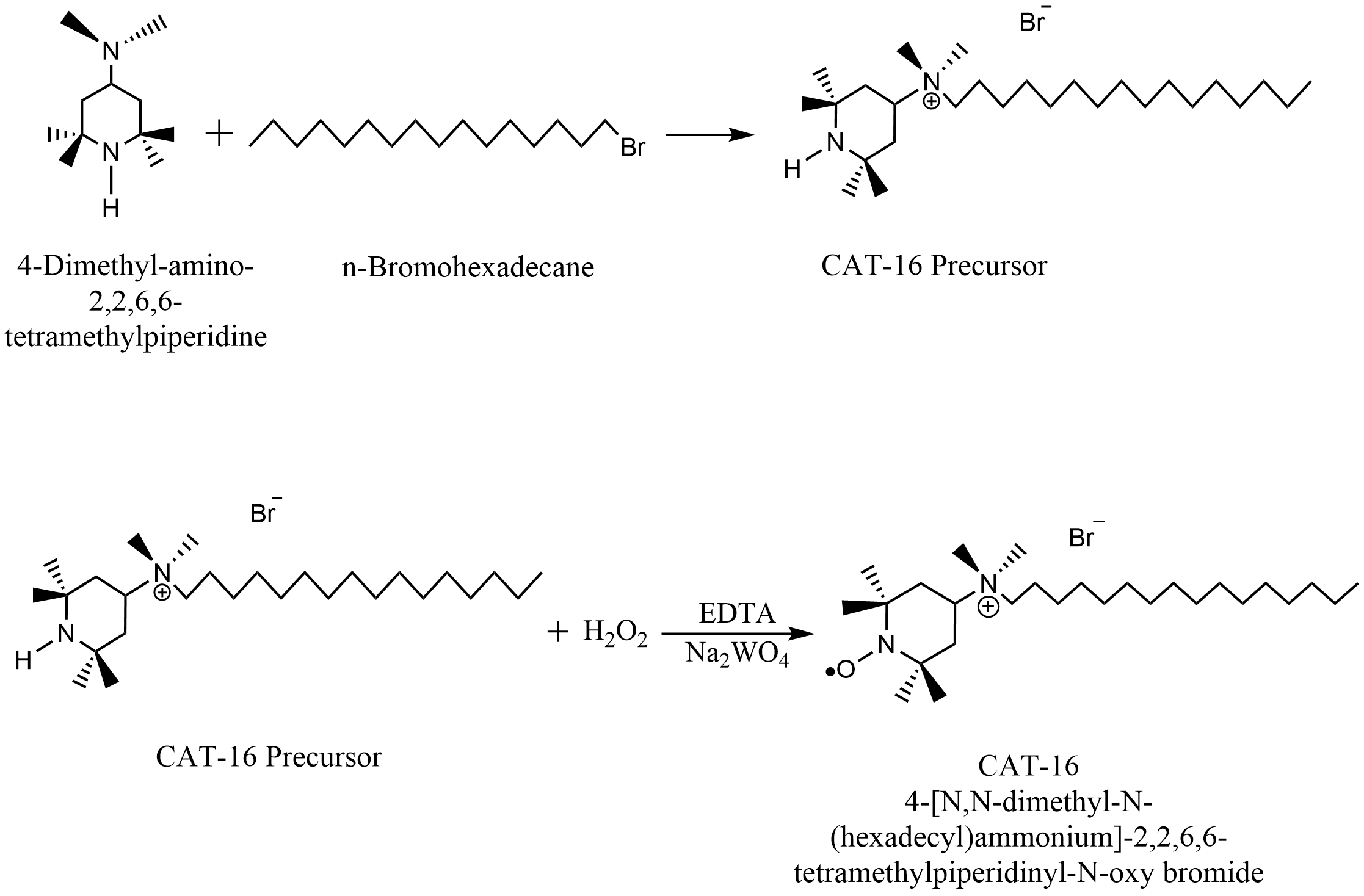}
\caption{Synthesis reaction of CAT-16}\label{fig:CAT16Reaction}
\end{figure}

The amphiphilic spin label CAT16
(4-{[}N,N-dimethyl-N-(hexadecyl)ammonium{]}-2,2,6,6-tetramethylpiperidinyl-N-oxy
bromide) incorporates into the surfactant-water interface of the \gls{rm}. A
protocol for the routine synthesis and purification of CAT16 was
developed based on previous literature
\cite{Haering1988ChaEleSpi,Angelov2014EPRRheStu,Hubbell1970IntSmaMol}.
It involves two steps: alkylation and oxidation.

\emph{Alkylation:} 1 mL (3.34 mmol) of
4-dimethyl-amino-2,2,6,6-tetramethylpiperidine (Santa Cruz
Biotechnology) was added to 1.25 mL (4.06 mmol) of bromohexadecane
(98\%, Alfa Aesar) in a round bottom flask. The mixture was heated in an
oil bath at 75°C for 24 hours under vigorous stirring in the dark, under
nitrogen flow. Following trituration with 15 mL hexane (95\%, Aldrich
Chemical Company) and vacuum filtration, the resulting solid retentate
was washed three times with 10 mL diethyl ether ($\geq 99.8\%$,
Sigma-Aldrich) to remove any side products, yielding 0.392 g (0.8 mmol)
white powder labeled as ``CAT-16 precursor''
(fig.~\ref{fig:CAT16Reaction}, 4-Piperidinaminium,
N-hexadecyl-N,N,2,2,6,6-hexamethyl bromide).

\emph{Oxidation:} The CAT-16 precursor (0.8 mmol, 0.392 g) was dissolved
in 5 mL methanol (HPLC grade, Fisher Chemical) to which 0.5 mL (2.21
mmol) 15\% H\textsubscript{2}O\textsubscript{2} (Macron Fine Chemicals)
was added dropwise in the presence of 24 mg (0.07 mmol) sodium tungstate
oxide dihydrate ($\geq 95\%$, Alfa Aesar) and 23 mg (0.06 mmol) EDTA
disodium salt dihydrate (99\%, grade, Fisher Chemical), while the
solution pH was maintained at 10 (with NaOH, Fisher Scientific). The
concentration of H\textsubscript{2}O\textsubscript{2} was verified using
peroxide test strips from Bartovation (Item model number: PPX01V50),
which are sensitive up to 10\% H\textsubscript{2}O\textsubscript{2}
(dilution allows accurate measurement of higher concentrations). The
reaction progress was monitored each day with TLC (solvent:
CHCl\textsubscript{3}:MeOH:H\textsubscript{2}O 13:5:1 v:v:v);
visualization: combined I\textsubscript{2} and UV; Chloroform: ACS grade
$\geq 99.8\%$, Sigma-Aldrich; Methanol: HPLC grade, Fisher chemical).
Additional H\textsubscript{2}O\textsubscript{2} was added in 0.25 mL
(1.1 mmol) increments until completion of reaction while maintaining pH
10. In total 1.5 mL (6.62 mmol) H\textsubscript{2}O\textsubscript{2} was
added over four days, and a mild yellow/brown colored liquid oil was
formed. (Note that the amounts and concentration of
H\textsubscript{2}O\textsubscript{2} mentioned above are the actual
tested values.) On the fifth day, the reaction mixture was extracted
$2\times$ with 20 mL dichloromethane (HPLC grade, Thermo Scientific)
and dried over anhydrous magnesium sulfate (Bio Basic). After rotary
evaporation, a reddish-orange thick oil remained; this solidified after
two hours when placed in a desiccator under vacuum. After $2\times$
trituration with 30 mL hexane, centrifugation, and multiple
recrystallizations from a mixture of dichloromethane and hexane, pure
salmon-colored CAT-16 was obtained. Dissolution in DMSO-d6 (99.9\%,
Cambridge Isotope Laboratories) with a drop of phenyl hydrazine (97\%,
Acros Organics) enabled \gls{nmr} at 800 MHz (fig.~\ref{fig:NMR_CAT16}):
\textsuperscript{1}H \gls{nmr} (DMSO-d\textsubscript{6}, quenched with phenyl
hydrazine, 800 MHz): $\delta$ 0.86 (t, 3H), 1.08-1.14 (d, 12H), 1.25
(m, 29H), 1.63 (m, 4H), 1.94 (d, 2H), 2.96 (s, 6H), 3.23 (m, 2H), 3.58
(t, 1H). \gls{esr} of CAT-16 dissolved in 90:10 v:v Chloroform:Methanol
confirmed the presence of the nitroxide group.

\subsection{\gls{rm} preparation}

A stock solution of 100 mM TEMPO-SO\textsubscript{4} (0.498 mmol, 0.145
g, fig.~\ref{fig:TEMPOSO4Reaction}) in 5 mL ultrapure water (Millipore
Milli-Q system) was prepared. Additionally, a stock solution of 2.03 mM
CAT-16 (0.204 mmol, 0.103 g, fig.~\ref{fig:CAT16Reaction}) in 100 mL
chloroform:methanol (90:10) (Chloroform: ACS grade $\geq 99.8\%$,
Sigma-Aldrich, Methanol: HPLC grade, Fisher Chemical) was prepared. The
TEMPO-SO\textsubscript{4} solution was diluted with ultrapure water to
the desired concentration for individual preparations. Stock solution in
CHCl\textsubscript{3}:MeOH solvent were stored at -20°C, while aqueous
stocks were stored at 4°C. Next, a solution of 0.300 M \gls{aot} surfactant
(Dioctyl sulfosuccinate sodium salt, $\geq 97\%$, Sigma-Aldrich, 1.50
mmol, 0.667 g) was prepared in 5 mL dispersant in either isooctane
(2,2,4-Trimethylpentane, Alfa Aesar) or in dodecane (n-Dodecane
$\geq99.8\%$, Thermo Scientific).

To prepare 200 μL \gls{rm} with CAT-16, first, the desired amount of stock
solution (typically 49.2 μL yielding $100 \times 10^{-9}$ mol CAT-16)
was dried in a dry culture tube under a flow of dry nitrogen before
adding 100 μL of 0.300 M \gls{aot} in dispersant and vortexing. Alternately,
for the TEMPO-SO\textsubscript{4} \gls{rm}, 100 μL of 0.300 M \gls{aot} (30.0 μmol)
in dispersant was added to a clean, dry culture tube (without CAT16). In
either case, to achieve the required water loading
($w_0 = \text{mol H}_2\text{O}/\text{mol \gls{aot}}\); \(3 \le w_0 \le 40$),
between 1.62 μL (90.0 μmol H\textsubscript{2}O) and 21.6 μL (1.20 mmol
H\textsubscript{2}O) of aqueous solution was added. (Note that the ratio
of the moles of water molecules to the aqueous solution volume was
assumed to be unchanged, at 55.5~M, for the TEMPO-SO\textsubscript{4}
solution).

Finally, the solution was diluted with pure dispersant to a total volume
of 200 μL, then vortexed (for repeated $60\;\text{s}$ intervals,
continuing until transparent -- typically 15 min). In all instances, the
concentration of moles of \gls{aot} per total \gls{rm} solution was maintained at
150 mM. The \gls{rm} samples equilibrated at room temperature for 30 minutes.
Each 20.0 μL sample was loaded into a glass capillary tube (0.704 mm
i.d. 1.20 mm o.d.) and held inside the end of a larger capillary so that
it was directly exposed to the flow of cold gas in the center of the
cavity.

\section{Results and Discussion}

\subsection{Heisenberg Exchange Responds to Segregation of the AqueousSolution}

\begin{figure}
\centering
\includegraphics[width=\linewidth]{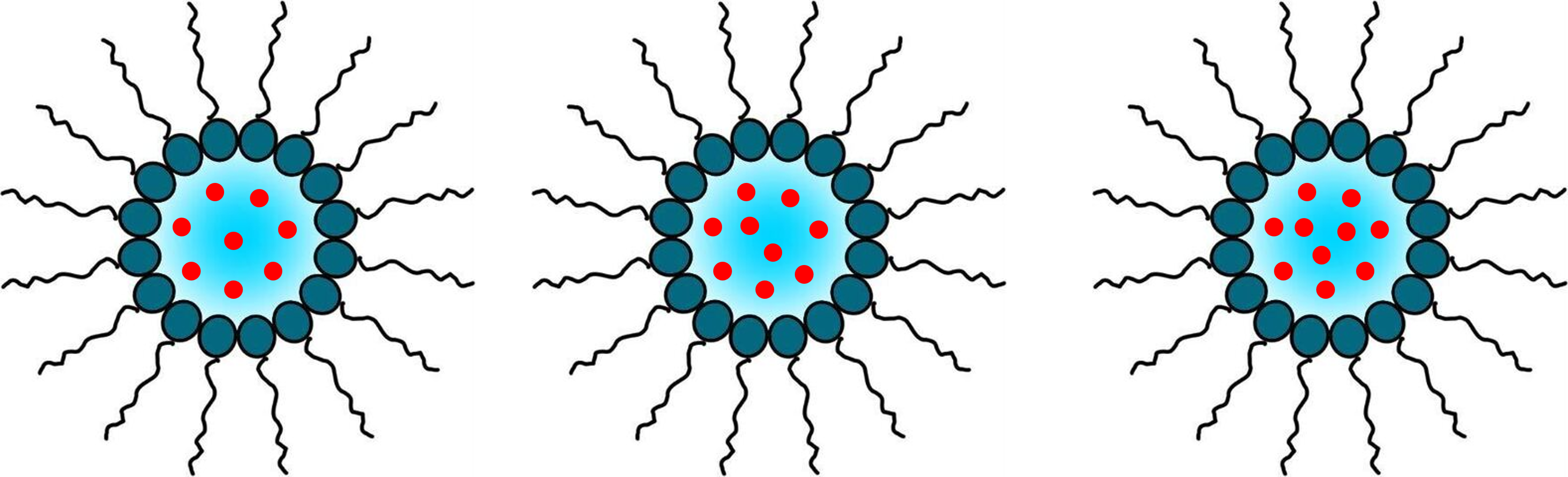}
\caption{When loaded with aqueous solution of spin probes (shown in red
here, in numbers appropriate for $w_0=20$ and local concentration of
100 mM) \glspl{rm} aggregates exhibit a range of local concentrations, from low
(left) to high (right). The different local concentrations result in
signals with different Heisenberg exchange rates.}\label{fig:HEcartoon}
\end{figure}

First, briefly note that, at high spin probe concentration, Heisenberg
Exchange leaves a distinctive signature on the spectra arising from the
segregation of the aqueous solution into the nanocomparments. For
example, very high concentration TEMPO-sulfate spin probes loaded into
\glspl{rm} yield a spectrum that changes significantly upon changes in water
loading ($w_0$), as shown in fig.~\ref{fig:HEiso} and
fig.~\ref{fig:HEdod}. These changes arise from differing distributions
of local concentrations inside the \gls{rm} at the two different $w_0$.
Specifically, subdivision of the aqueous solution into nanocompartments
yields a variation of the local concentration inside each
nanocompartment, since even at high concentration, each \gls{rm}
compartment/aggregate will include, at most, tens of individual spin
probe molecules (fig.~\ref{fig:HEcartoon}). In fact, a simple Poisson
distribution analysis (fig.~\ref{fig:poisson}) suggests that when a 100
mM spin probe solution is added to the surfactant to create an \gls{rm} of
$w_0=20$, the local concentration of spin probe inside the \gls{rm}
frequently varies significantly from the average concentration. Each
different local concentration results in a differing rate of spin probe
collisions, Heisenberg exchange, and therefore \gls{esr} linewidth.

\begin{figure}
\centering
\includegraphics[width=3.5in]{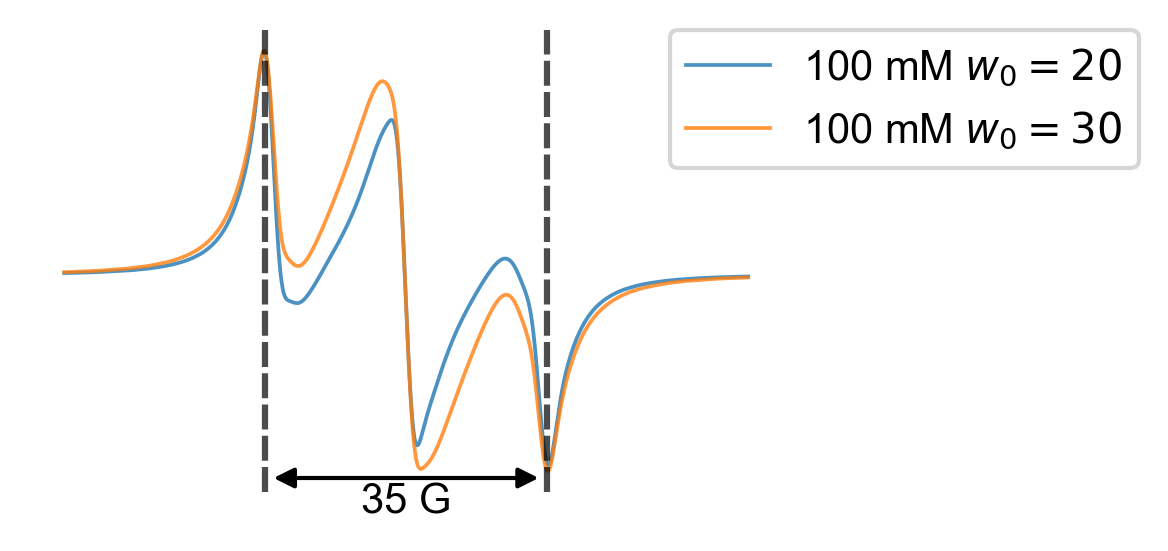}
\caption{Room temperature \gls{esr} (peak-normalized) of 100 mM
TEMPO-SO\textsubscript{4} in isooctane dispersant at $w_0=20$ (blue)
and $w_0=30$ (orange).}\label{fig:HEiso}
\end{figure}

\begin{figure}
\centering
\includegraphics[width=3.5in]{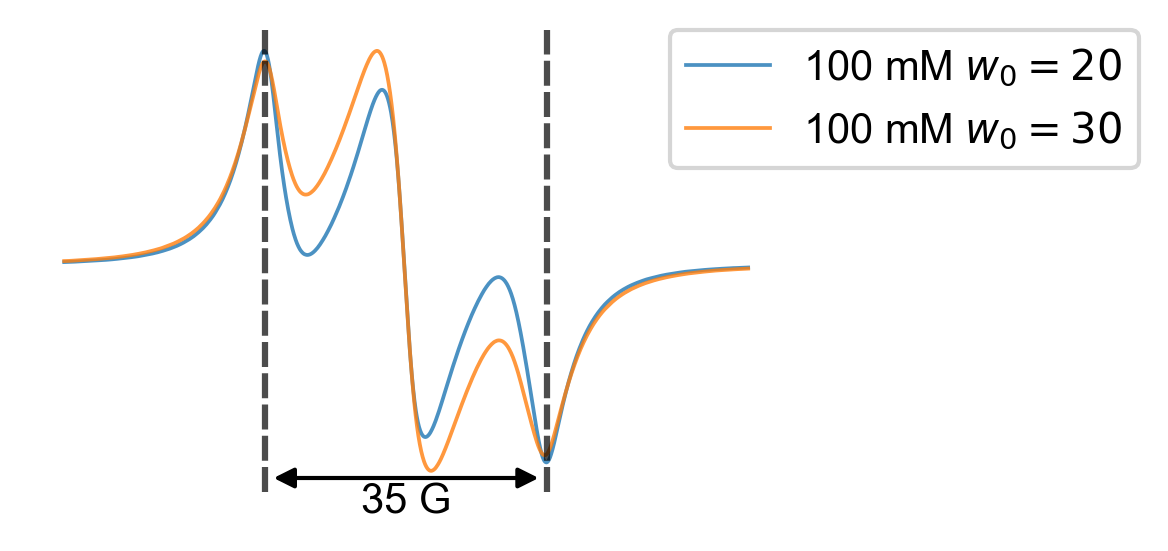}
\caption{Room temperature \gls{esr} (peak-normalized) of 100 mM
TEMPO-SO\textsubscript{4} in dodecane dispersant at $w_0=20$ (blue)
and $w_0=30$ (orange).}\label{fig:HEdod}
\end{figure}

In particular, note that individual \glspl{rm} with high local concentration
(near to 100 mM and greater) and correspondingly rapid Heisenberg
exchange contribute to the broad central hump in fig.~\ref{fig:HEdod}
and fig.~\ref{fig:HEiso}. When some population of \glspl{rm} with lower local
concentrations are present, they contribute spectra that have slower
Heisenberg exchange and are better resolved into three individual
hyperfine components. Because these contributions have a smaller
linewidth and because the amplitude of the derivative signal is
inversely proportional to the square of the linewidth, the spectrum
responds correspondingly dramatically to contributions from the \glspl{rm} with
low local concentration. The Poisson distribution predicts that lower
$w_0$ \gls{rm} mixtures will contain significantly more \gls{rm} aggregates with a
low local concentration (consider the number of \glspl{rm} with 25 mM or 50 mM
in fig.~\ref{fig:poisson}). In contrast, the distribution of local
concentrations inside \glspl{rm} with large $w_0\) (\(w_0=30$ in
fig.~\ref{fig:poisson}) peaks more sharply about a single concentration.
Correspondingly, the resolved hyperfine components of the spectra
contribute a smaller fraction of the overall spectrum
(fig.~\ref{fig:HEdod} and fig.~\ref{fig:HEiso}).

\begin{figure}
\centering
\includegraphics[width=3.5in]{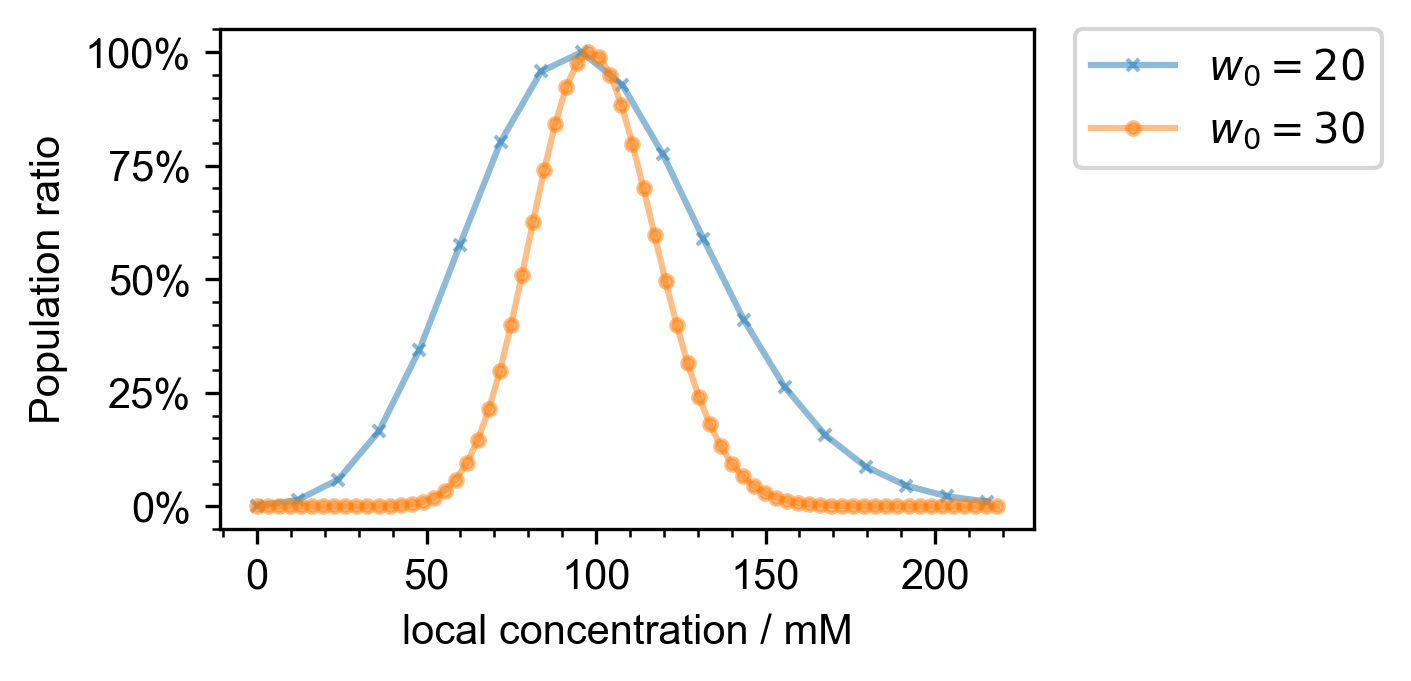}
\caption{Predicted percentage of \glspl{rm} with each local concentration,
normalized against population of most likely concentration. Note in
particular how, at lower $w_0$ (blue line), the low-concentration
portion of the distribution includes a significant number of \glspl{rm} with
lower concentration. At higher $w_0$, the distribution is predicted to
peak more tightly about the concentration of the solution before it was
added to the \glspl{rm}.}\label{fig:poisson}
\end{figure}

In fig.~\ref{fig:poisson} a Poisson distribution preddicts the
distribution of local concetrations based on the average number of water
molecules present in the \gls{rm}. Specifically, the formula in
\cite{Beaton2023RapScrCor} (which is derived from a meta-analysis of
the literature) demonstrates that a \gls{rm} with $w_0=20$ contains 4640
water molecules, so that the ratio of moles of \gls{rm} aggregates per total
volume of water (or aqueous solution) in the \gls{rm} solution is 11.9~mM
while a \gls{rm} with $w_0=30$, due to its larger diameter of 14.75 nm,
contains 17,100 water molecules and thus exhibits an aggregate to water
volume ratio of only 3.25~mM. This \gls{rm} aggregate:water ratio proves
useful, because when the aqueous concentration of a guest molecule (such
as the spin probe here) is divided by this number, it gives the average
number of guest molecules per \gls{rm} aggregate. For example, at an aqueous
concentration of 100 mM spin probe, $w_0=30$ \gls{rm} aggregates contain an
average of 30.7 spin probe molecules while $w_0=20$ \gls{rm} aggregates
contain only 8.35 spin probes.

Further investigation of Heisenberg exchange in \glspl{rm} offers promise for
informing on the dynamic interactions and fusion events between these
nanocontainers. In particular, the differences in the
fig.~\ref{fig:HEdod} and fig.~\ref{fig:HEiso} likely arise due to
differing rates of fusion/fission events, which mix the local
concentrations and could therefore affect the Heisenberg exchange rates.
However, these considerations are reserved for future study and the
following studies employ a concentration of 20~mM or less to avoid such
complications.

\subsection{Dramatic Variation of Spin Label Mobility with Temperature}

Even a standard liquid nitrogen temperature control unit can achieve a
wide enough range of temperatures to dramatically change the mobility of
a spin probe inside a \gls{rm}. In fact, the studies presented here explore
almost the full dynamic range of standard X-band \gls{esr} experiments, as the
line shapes vary from nearly isotropic three-line spectra, consistent
with a spin probe freely dissolved in a viscous liquid, to spectra that
are broad enough to be consistent with glassy solids. As an example,
fig.~\ref{fig:justSpectra} shows the lineshape variation exhibited by
TEMPO-SO\textsubscript{4} inside a \gls{rm} (dispersed in isooctane) that is
small enough that water inside does not freeze (and thus exhibits no
abrupt discontinuities of lineshape). Fig.~\ref{fig:justSpectraCat16}
illustrates a similar variation for a CAT-16 spin label.

\begin{figure}
\centering
\includegraphics[width=3in]{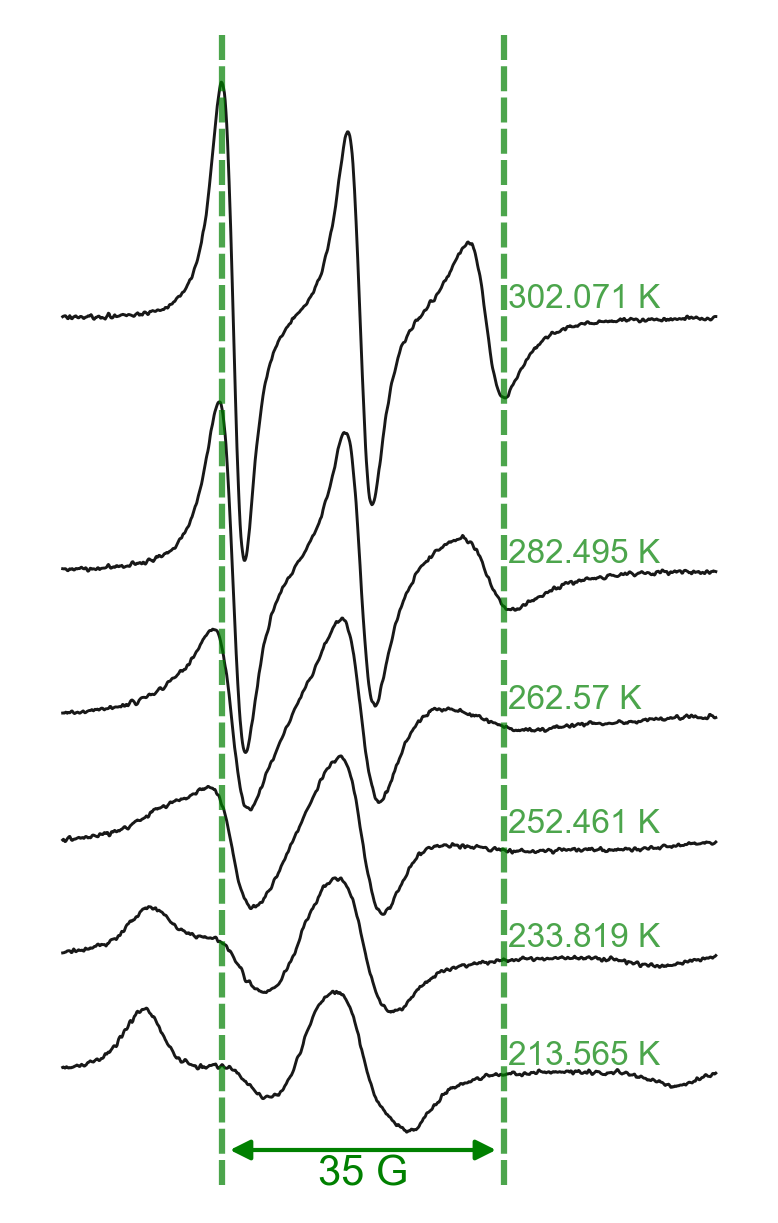}
\caption{\gls{esr} spectra (not normalized) of TEMPO-SO\textsubscript{4} in
$w_0=3$ \glspl{rm} in isooctane over a range of temperatures demonstrate an
almost complete exploration of the dynamic range of X-band \gls{esr}
spectroscopy.}\label{fig:justSpectra}
\end{figure}

\begin{figure}
\centering
\includegraphics[width=3in]{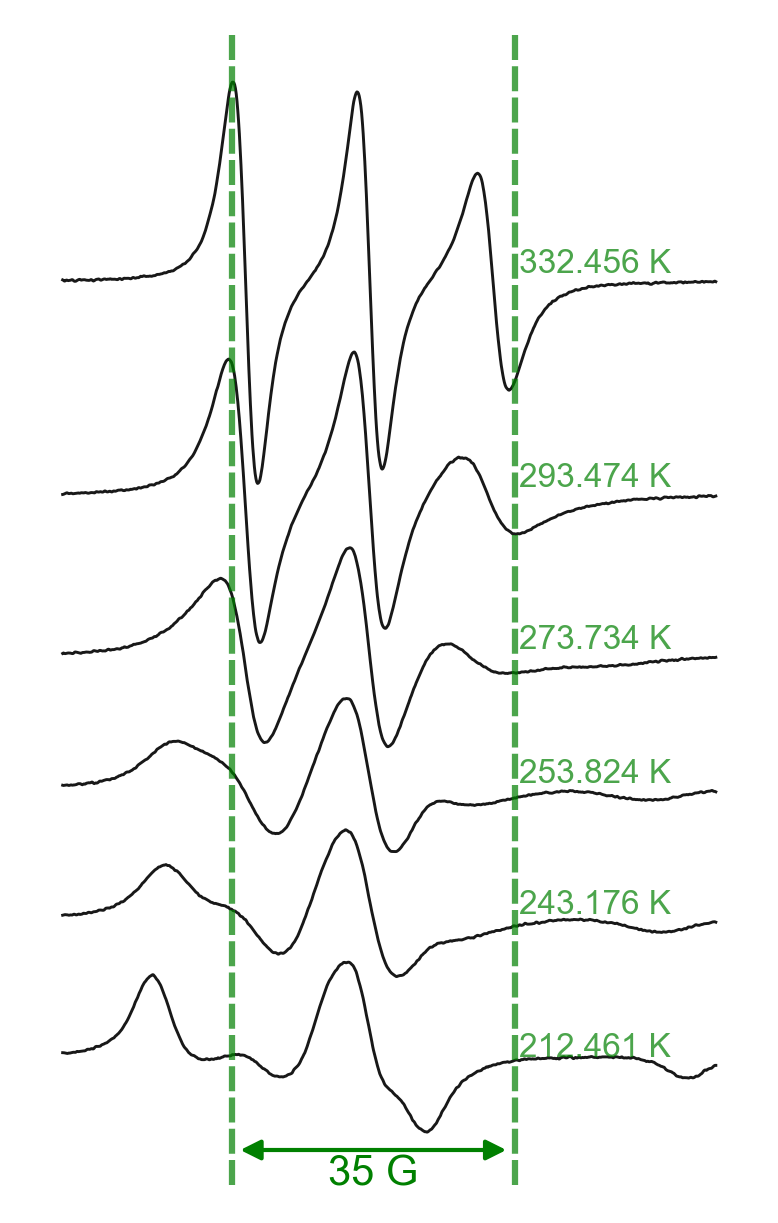}
\caption{\gls{esr} spectra (not normalized) of CAT-16 in $w_0=3$ \glspl{rm} in
isooctane over a range of temperatures also demonstrate an almost
complete exploration of the dynamic range of X-band \gls{esr}
spectroscopy.}\label{fig:justSpectraCat16}
\end{figure}

Existing Python libraries enable rapid development of a script that
estimates the correlation time of these spectra. This script determines
the field positions and amplitudes of the peaks for an entire
variable-temperature experiment. It implements spline-based smoothing of
the spectra in order to minimize the impact of any noise as part of this
determination, making it relatively precise. Finally, it estimates the
rotational correlation time ($\tau_c$) \emph{via} the classic equation
\cite{Kivelson1960TheESRLin,Avramiotis1999IntProLec}:
\begin{equation}\tau_c \approx
    \left(6.51\times 10^{-10} \left( \frac{\text{s}}{\text{G}} \right)\right)
    \Delta B_0
    \left[
        \sqrt{\frac{h_0}{h_{+1}}}
        +
        \sqrt{\frac{h_0}{h_{-1}}}
    -2 \right]
\label{eq:tauFromLinewidth}\end{equation} where $\Delta B_0$ gives
the linewidth of the central line, and $h_{m}$ gives the peak-to-peak
height for the hyperfine transition associated with nuclear quantum
number $m\) (the three peaks \(m=+1, 0, -1$ appear from low to high
field, respectively). (Note that here and elsewhere, units associated
with numbers are given in equations in Roman font -- \emph{e.g.}
$\text{G}\), \(\text{s}$, \emph{etc.})

This approximate measure of correlation time enables simple
visualization of the change in the \gls{esr} spectra as a function of
temperature. While it provides less, and more approximate, information
\emph{vs.} a simulation, it also doesn't require rigorous cross-checking
of the various simulated parameters. Furthermore, it will focus on the
most intense component of a multi-component spectrum. Strictly,
eq.~\ref{eq:tauFromLinewidth} strictly only accurately estimates the
correlation time in the fast-motion regime,
\cite{Avramiotis1999IntProLec} \emph{i.e.}
$10^{-11}<\tau_c< 3\times 10^{-9}\;\text{s}$. Nonetheless, it will
also be shown that, even outside this range of applicability,
eq.~\ref{eq:tauFromLinewidth} provides a simple single number sensitive
to dramatic changes in the lineshape (or lack thereof).

By exploring molecular motion over a wide temperature range, \gls{esr} can
offer insight into the unusual properties of water under confinement
and, in particular, how it varies with temperature. In a system such as
\glspl{rm}, the diffusivity of molecules varies from site to site. As a result,
thermal energy excites the spin probe to different rates of rotation
when it resides in a fluid aqueous medium at the core of the \gls{rm},
\emph{vs.} in the interfacial layer next to the surfactant, \emph{vs.}
in very small \glspl{rm} where relatively immobile water and surfactants might
rotate as a single unit inside the dispersant medium. Furthermore, in
all these cases, the thermal energy needed to activate diffusion of the
molecules surrounding the spin probe (water, interfacial water, or
dispersant, depending on context), and thus to reduce resistance to such
rotation, varies. As a result, the spectrum of the spin label rotating
inside these media will respond to temperature differently. Thus,
observation of the temperature dependence of the rotational correlation
time offers important insight into the energy barriers associated with
transport inside nanoscopically heterogeneous materials and, in
particular, offers a unique opportunity to characterize the energies
required to move solvent molecules from site to site within the \gls{rm}
solution.

Thus, this contribution expands on the well-documented capability of \gls{esr}
to measure microviscosity
\cite{Fukuda2001,Lewinska2012ChaMicAlk,Lawton2010}, and focuses on
how the mobility (specifically the diffusivity) of the confined water,
the dispersant, or both, responds to changes in temperature. In
particular, the combination of several standard relations enables the
quick characterization of the variable temperature spectra in terms of
activation energies for diffusion. To begin, \gls{sde}:
\begin{equation}\tau_c= \frac{4\pi\eta r^3}{3 k_B T}\label{eq:SDEeq}\end{equation}
relates the correlation time ($\tau_c$) of
eq.~\ref{eq:tauFromLinewidth}, to the viscosity of the solvent
($\eta$, either water or dispersant, depending on context), where
$k_B\) is Boltzmann's constant, \(T\) is the temperature, and \(r$ is
the radius of the species undergoing rotational diffusion (either the \gls{rm}
or the spin probe itself, depending on context). Next, to relate the
translational diffusivity ($D$) of a particle to the viscosity of the
surrounding solvent, the Einstein relation
\[\eta = \frac{k_B T}{6\pi D r_{solv}},\] gives an expression for the
viscosity of the solvent ($\eta$) in terms of its self-diffusivity
($D\)), the radius of the solvent molecules (\(r_{solv}$), and the
temperature. Substitution of this expression into eq.~\ref{eq:SDEeq} gives
an expression showing that the correlation time is inversely
proportional to the self-diffusivity of the solvent
\[\tau_c = \frac{2r^3}{9Dr_{solv}}.\] Because the rotational diffusion
quantified by $\tau_c$ and translational solvent self-diffusion
quantified by $D$ are both driven by thermal fluctuations, the
$k_BT$ dependence has canceled out. The assumption that the solvent
self-diffusivity obeys an Arrhenius dependence
($D\approx A\exp(-E_a/RT)$, over a limited range of temperatures) then
yields \[\tau_c = \frac{2r^3}{9Ar_{solv}}e^{+\frac{E_a}{RT}},\] and the
logarithm of both sides yields
\begin{equation}\protect\hypertarget{eq:arrhenius}{}{\log(\tau_c) = 
const+\frac{E_a}{2.3026 RT},}\label{eq:arrhenius}\end{equation} where
$\tau_c\) is the correlation time, \(T\) is the temperature, \(E_a$ is
the activation energy associated with the diffusivity of the solvent,
$R\) is the ideal gas constant, and \(const$ is a constant involving a
collection of terms (including the Arrhenius prefactor, $A$,
hydrodynamic radii, \emph{etc.}) Thus, the left-hand side of
eq.~\ref{eq:arrhenius} plotted against $1000/T$ yields a classic
Arrhenius plot for the diffusivity of the solvent.

Such an Arrhenius plot of the estimated correlation time
(eq.~\ref{eq:tauFromLinewidth}) experienced by a
TEMPO-SO\textsubscript{4} inside the aqueous core of an \gls{rm} dispersed in
isooctane, in fig.~\ref{fig:ArrhTEMPOIsooctane}, yields several
unexpected results. To interpret these results, first note that one can
assume that $\tau_c$ in eq.~\ref{eq:arrhenius} decomposes into
\begin{equation}\protect\hypertarget{eq:recipAdd}{}{
\tau_c^{-1}
=
\tau_{\gls{rm}}^{-1}+\tau_{SP}^{-1}
}\label{eq:recipAdd}\end{equation} where $\tau_c$ gives the
correlation time observed by \gls{esr}, $\tau_{\gls{rm}}$ represents the
correlation time associated with the rotational diffusion of the entire
\gls{rm} aggregate (\emph{i.e} assembly, \emph{i.e.} nanocontainer) and
$\tau_{SP}$ represents the correlation time for the rotational
diffusion of the spin probe inside the water nanopool contained within
the \gls{rm} -- more concretely, the rotational diffusion relative to the
diffusion (or moment of inertia) frame of the \gls{rm} aggregate. Importantly,
when considering eq.~\ref{eq:SDEeq} and eq.~\ref{eq:arrhenius}, the
hydrodynamic radius of the spin probe and the diffusivity (viscosity) of
water control $\tau_{SP}$; in contrast, the hydrodynamic radius of the
\gls{rm} aggregate and the diffusivity (viscosity) of the dispersant control
$\tau_{\gls{rm}}$.

Note that $w_0$ controls the size of the \gls{rm} aggregates: specifically,
$r\approx (0.175\;\text{nm})w_0+c\), where the constant \(c$ depends
on the thickness of the surfactant
\cite{maitraDetermination1984,luisiReverse1988}. Therefore, \glspl{rm} at
the limit of low \emph{vs.} high $w_0$ will approach different
limiting cases of eq.~\ref{eq:recipAdd}. The limit
$\tau_c \approx \tau_{\gls{rm}}$, is most likely to occur for small \gls{rm} sizes
(small $w_0$) and would specifically correspond to spin labels that
were fixed inside \gls{rm} aggregates undergoing rotational diffusion, with
the $r$ in eq.~\ref{eq:SDEeq} corresponding to the radius of the \gls{rm}
aggregate and $E_a$ the activation energy for the diffusivity of the
dispersant. In contrast, the limit $\tau_c\approx\tau_{SP}$ would
correspond to large $w_0$, as the larger \glspl{rm} likely contain less
viscous water and also exhibit slower tumbling of the \gls{rm} aggregate due
to their larger radius, with the $r$ in eq.~\ref{eq:arrhenius}
corresponding to the hydrodynamic radius of the spin probe and $E_a$
corresponding to the activation energy for the diffusion of water inside
the \gls{rm}.

\begin{figure}
\centering
\includegraphics[width=\linewidth]{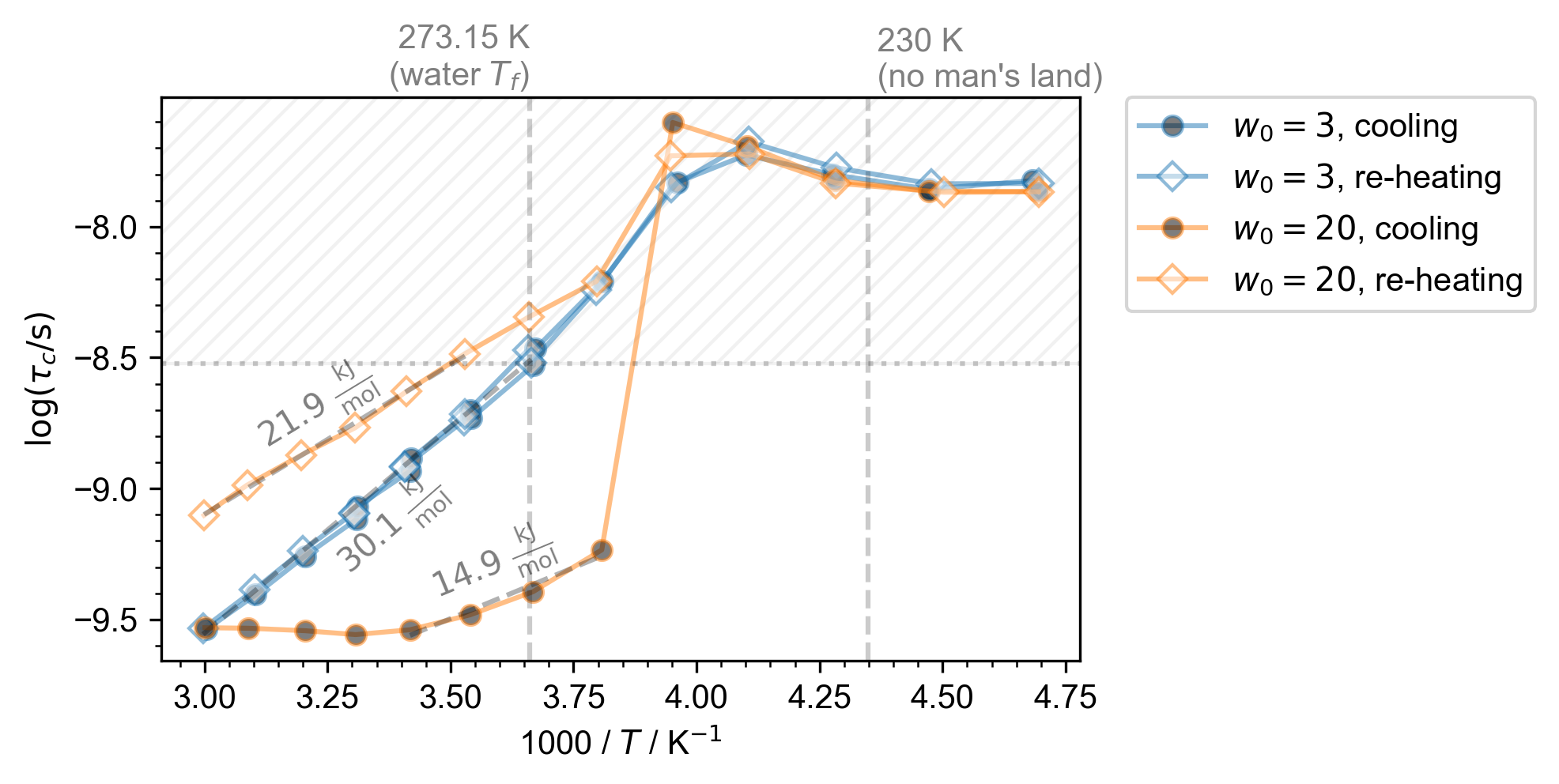}
\caption{Arrhenius temperature dependance for the rotational correlation
time of 20 mM TEMPO-SO\textsubscript{4} in isooctane. Note that
datapoints representing slower motion appear further up along the
$y-$axis and that datapoints representing colder temperatures appear
further to the right. The crosshatched region indicates the regime over
which eq.~\ref{eq:tauFromLinewidth} is not expected to hold
quantiatively. While the correlation time of the smaller $w_0=3$ \gls{rm}
upon cooling (dark blue circles) indicates Arrhenius diffusivity of the
solvent and is reproduced upon subsequent reheating of the same sample
(blue diamonds), cooling of larger ($w_0=20$) \glspl{rm} (dark gold circles)
results in an abrupt transition slightly below the freezing point of
water. Furthermore, the cold-shedding of the water alters the \glspl{rm} so
that the lineshapes upon re-heating (gold diamonds) differ
significantly, as discussed in the text.}\label{fig:ArrhTEMPOIsooctane}
\end{figure}

If the $\tau_c\approx\tau_{\gls{rm}}$ limit held strictly true here for the
smaller \gls{rm} aggregates ($w_0=3$), the spin probe would experience a
$\tau_c\approx \tau_{\gls{rm}}\) of \(4.00\ \mathrm{ns}$ (\emph{i.e.}
$\log(\tau_c/\text{s})=-8.398$), from eq.~\ref{eq:SDEeq}, given that the
viscosity of isooctane is \cite{Padua1996DenVisMea}
$0.473\ \mathrm{mPa}\cdot\mathrm{s}$ at room temperature and the \gls{rm}
radius at $w_0=3\) is approximately \(2.02\ \mathrm{nm}$. The faster
rotational correlation time reported in
fig.~\ref{fig:ArrhTEMPOIsooctane} (blue) thus reflects a spin probe not
totally immobilized inside the \gls{rm}, likely due in part to lubrication by
the water molecules. In fact, assuming the calculated
$\tau_{\gls{rm}}=4.00\;\text{ns}$ here, and subtituting the measured
$\tau_c=1.02\;\text{ns}$ (\emph{i.e.}
$\log(\tau_c/\text{s})=-8.990$, blue diamonds in
fig.~\ref{fig:ArrhTEMPOIsooctane}) into eq.~\ref{eq:recipAdd} yields a
value of $\tau_{SP}\approx 1.37\;\text{ns}$, indicating that the
motion of the spin label inside the \gls{rm} likely provides the primary
mechanism for rotational diffusion. The slopes of the line for both the
cooling of the $w_0=3$ \gls{rm} (dark blue circles), and the re-heating of
same sample (blue diamonds), indicate an activation energy of
$E_a= 29.4\;\text{kJ/mol}$. This value significantly exceeds the
$E_a$ of diffusivity of isooctane (calculated by fitting an Arrhenius
plot for the literature values \cite{Padua1996DenVisMeaa} of
$T/\eta\) of to a straight line), which is \(11.2\;\text{kJ/mol}$.
Thus, this likely reflects the added energy involved in moving the
lubricating water molecules that surround the spin probe in order to
achieve the faster than expected correlation time.

A \gls{rm} with $w_0=20$, in contrast, has a radius of
$5.35\ \mathrm{nm}$, corresponding to
$\tau_{\gls{rm}}\approx 73.7\;\mathrm{ns}$ (\emph{via} eq.~\ref{eq:SDEeq}) in
isooctane dispersant at room temperature. Comparison of this value to
fig.~\ref{fig:ArrhTEMPOIsooctane} (and keeping in mind the reciprocal
addition of eq.~\ref{eq:recipAdd}) indicates that the tumbling of the \gls{rm}
is not a significant contributor to the $\tau_c$ experienced by the
spin probe. Therefore, it is sensible to consider the limit
$\tau_c\approx\tau_{SP}$, estimated with a hydrodynamic radius of
$0.34\;\mathrm{nm}$ (\emph{i.e.}, close to that of TEMPOL
\cite{Banerjee2009ESREviCoe}) for TEMPO-SO\textsubscript{4} and the
standard viscosity of water at room temperature in order to predict a
$\tau_{SP}$. These assumptions yield \(\tau_c= 35.6\
\mathrm{ps}\) (a similar order of magnitude as
$\tau_c= 15\;\mathrm{ps}$ for TEMPOL in water as noted in other
literature \cite{Peric2013RotFouSmaa}), which is almost an order of
magnitude faster than the $\tau_c = 280\;\text{ps}$ (\emph{i.e.}
$\log(\tau_c/\text{s})=-9.55$, orange diamonds in
fig.~\ref{fig:ArrhTEMPOIsooctane}) correlation time observed for the
spin probe tumbling inside the confined water. Thus, the
TEMPO-SO\textsubscript{4} spin probe experiences significantly greater
microviscosity than one would expect for water in bulk solution. Of
course, observation of such slow water under confinement or at
interfaces proves largely consistent with the literature.

The rotational correlation time of the spin probe in the $w_0=20$
sample changes very little upon cooling to slightly below the freezing
point of water. This lack of change over such a wide temperature range
is striking. It seems to imply a very small activation energy for the
diffusivity of water. However, water inside confined systems where
diffusion remains relatively unrestricted in one dimension (\emph{i.e.}
inside tubes or between bundles) generally exhibits a lower limit
activation energy of diffusion of 18.5~kJ/mol for bulk liquid water.
Under confinement, if anything, the activation energy for the diffusion,
and therefore viscosity of water, is expected to increase
\cite{Zhang2020RelGeoNan}.

\begin{figure}
\centering
\includegraphics[width=3.5in]{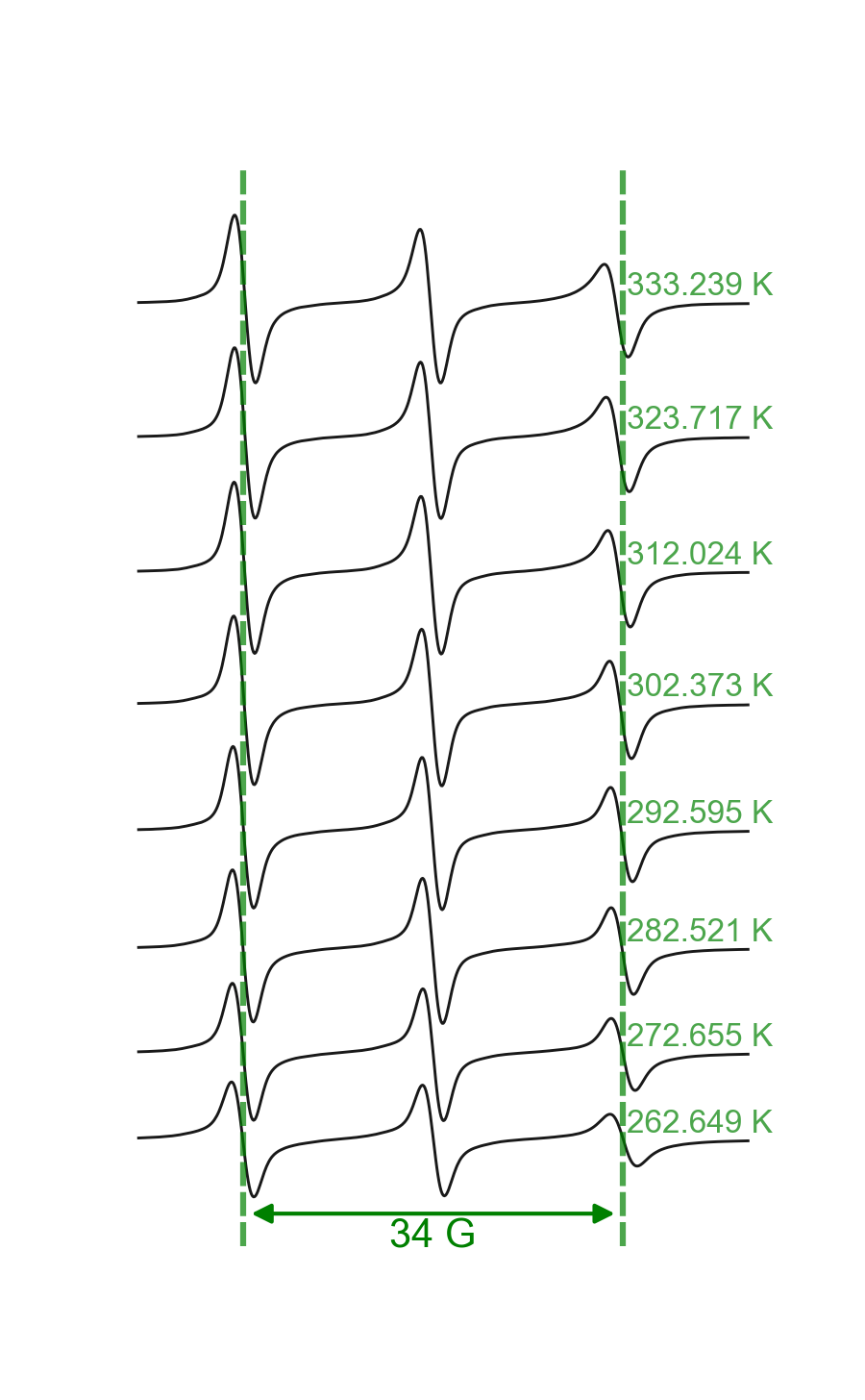}
\caption{\gls{esr} spectra of TEMPO-SO\textsubscript{4} in $w_0=20$ \glspl{rm} in
isooctane over a wide range of temperatures.}\label{fig:TEMPO_Iso_w20}
\end{figure}

Such an unusual result invites significant speculation. One might
speculate that the linewidth analysis method for assessing the
correlation time fails wildly here. However, the lineshapes
(fig.~\ref{fig:TEMPO_Iso_w20}) indeed change very little with
temperature. In fact, the linewidths and relative amplitudes of the
derivative peaks that reflect the molecular dynamics (see
eq.~\ref{eq:tauFromLinewidth}) change very little. Only the relative
field position of the three peaks change (notably the position of the
rightmost peaks in fig.~\ref{fig:TEMPO_Iso_w20}), in a fashion that
suggests a change to the isotropic hyperfine constant. A changing
isotropic hyperfine constant, in turn, indicates changing polarity in
the environment around the nitroxide group \cite{Zuev2003EffProSol}.
Thus, the only significant changes to the lineshape over this extended
temperature range point not to changes in spin label mobility, but
rather to one of two scenarios. In the first scenario, the spin label
moves from sampling more of the interfacial layer or even some of the
surfactant to remaining more firmly centered inside the \gls{rm}. In the
second scenario, the distribution of the spin probe location within the
\gls{rm} remains fixed, but the average hydrogen bonding strength of the water
molecules increases. Notably, either or both scenarios might actually
cause the apparent lack of change of mobility over such a large
temperature range (\emph{i.e.} the very small apparent $E_a$).
Consider, for the first scenario, the distribution of spin probe
positions. In the interfacial layer, the water ostensibly moves more
slowly than in the central core. If decreases in temperature localize
the distribution of spin probe positions more tightly to the more fluid
central core, the change of location could counteract the decreasing
average mobility of the water populations as the \gls{rm} is cooled. Consider,
for the second scenario, that \glspl{rm} are not only a pool of water, but one
bounded and kept in isolation by strong interactions between the
surfactant molecules \cite{EskiciSizSizAOT2016,Urano2019AerSurFora}.
Temperature-driven changes to the structure of the surfactant molecules
and their associated ions may help to stabilize the water structure as
the temperature decreases, resulting in less disruption to the native
hydrogen-bonding matrix of water, trading better surfactant-surfactant
interactions for more fluid motion. Consistent with such a scenario,
previous 2D \textsuperscript{2}H \gls{nmr} spectroscopy have revealed that
confined environments that tend to disrupt the natural mobility of water
also seem to disrupt the natural hydrogen bond network
\cite{Beaton2023RapScrCor}. Thus, one might propose that
rearrangements of the surfactant molecules that stabilize the hydrogen
bond network improve the fluidity of the water enough to partially
offset the decreasing temperature. Finally, before moving on from the
region with apparently low $E_a$ diffusivity, it is worth noting that
something similar to the non-\gls{sde} curvature observed here has been
observed previously \cite{Peric2013RotFouSmaa}, where it was
interpreted in terms of density fluctuations of water.

\subsection{Observation of Water ``Cold-Shedding''}

\begin{figure}
\centering
\includegraphics[width=3.5in]{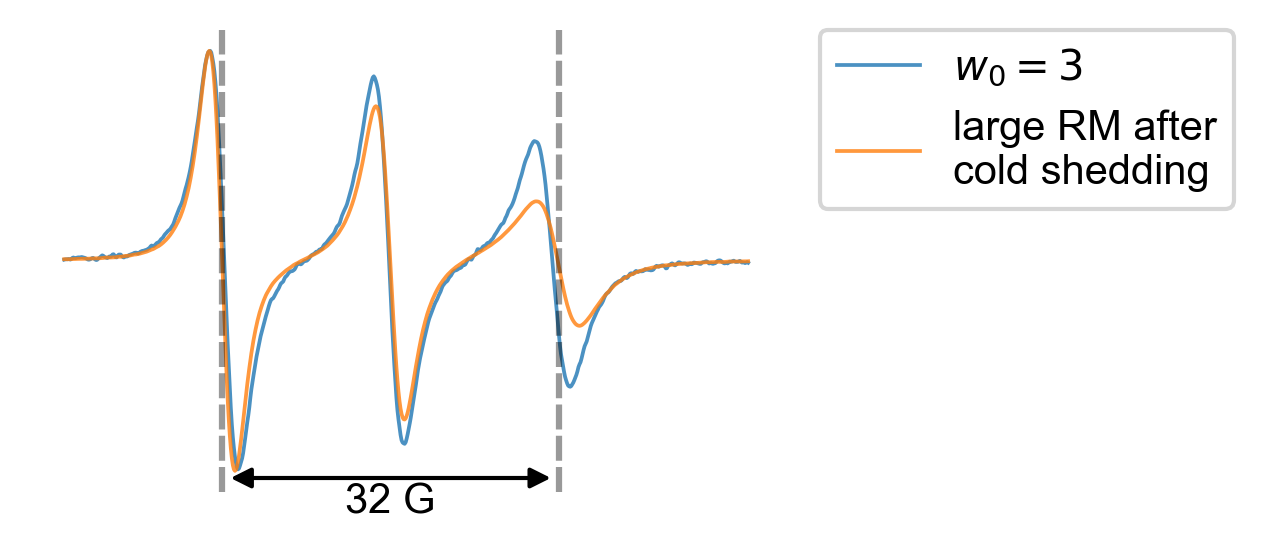}
\caption{\gls{esr} spectra (normalized) of 20 mM TEMPO-SO\textsubscript{4} in
$w_0=3\) and a large (\(w_0=20$) \gls{rm} in isooctane. Both have been
cycled below 213 K and back to 333 K temperature.}\label{fig:w3vsShed}
\end{figure}

In isooctane dispersant, once the temperature lowers slightly below the
freezing point of water, the line shape changes abruptly, giving rise to
the dramatic jump in fig.~\ref{fig:ArrhTEMPOIsooctane}. This transition
is caused by the ``cold shedding'' of water
\cite{Simorellis2006DynLowTem,VanHorn2009RevMicEnc}. The cold
shedding alters the sample so that the \gls{esr} of the sample after reheating
to room temperature appears consistent with a much smaller \gls{rm}.

Various details offer clues as to the identity of this ``post-shedding''
\gls{rm} (\emph{i.e.} the \gls{rm} that began as $w_0=20$, but which has changed
in composition after the temperature was cycled below the freezing point
and back to room temperature). As shown in fig.~\ref{fig:w3vsShed}, the
spectrum of the post-shedding \gls{rm} exhibits a significantly greater
variability in derivative peak amplitudes for the three hyperfine
transitions, giving rise (via eq.~\ref{eq:tauFromLinewidth}) to the
longer correlation time in fig.~\ref{fig:ArrhTEMPOIsooctane}.
Interestingly, the double integral of the EPR signal does not change
significantly over the course of the cycle from room temperature down to
cold temperatures and back to room temperature. This implies that the
spin probes all integrate into the new, smaller $w_0$ environment
rather than migrating to the end of the sample capillary with the water
that is shed as ice. This combined with the long correlation time
evidenced by fig.~\ref{fig:CAT16_ShedvsW0Series} indicate that the
temposulfate all remains inside the \glspl{rm} during the shedding process.
Also interestingly, the \gls{esr} spectrum for the $w_0=20$ \gls{rm} subject to
cold shedding (fig.~\ref{fig:w3vsShed}) actually exhibits a slower
correlation time than the $w_0=3$ \gls{rm}. Therefore, \gls{rm} that undergoes
shedding is certainly smaller than the $w_0=20$ \gls{rm}, since it doesn't
exhibit the fast correlation time associated with fluid water.
Interestingly, this smaller \gls{rm} also exhibits
(fig.~\ref{fig:ArrhTEMPOIsooctane}) a slightly lower activation energy
of diffusion than the $w_0=3$ \gls{rm} that likely indicates the
post-shedding \gls{rm} is slightly larger than the $w_0=3$ \gls{rm} and that the
water lubricating the spin probe rotation experiences slightly less
confinement than in the $w_0=3$ \gls{rm}, leading to a lower activation
energy.

Notably, the behavior of the (initially) $w_0=20$ \gls{rm} contrasts with
that of the $w_0=3$ \gls{rm}, which exhibits no evidence of cold shedding,
yielding the same spectra at the same temperatures during the reheating
process. Note that the correlation time of the post-shedding \gls{rm} may be
longer than that of the $w_0=3$ \gls{rm} for two reasons. In one possibility
the post-shedding \gls{rm} is smaller, and the low correlation time results
from less fluid water lubricating the spin probe. In another
possibility, the \gls{rm} that underwent shedding is larger and any increase
in water fluidity does not overcome the increase in $\tau_{\gls{rm}}$ in
eq.~\ref{eq:recipAdd}. A \gls{rm} of $w_0=7$ with an \gls{rm} radius of
$2.72\;\mathrm{nm}\) results in \(\tau_{\gls{rm}}\approx 9.74\;\mathrm{ns}$
(using eq.~\ref{eq:SDEeq}) in isooctane dispersant at room temperature.
Thus, $\tau_{\gls{rm}}^{-1}$ (\emph{i.e.} proportional to the \emph{rate} of
rotational diffusion) decreases by only $0.148\;\text{ns}^{-1}$ by
expanding from a $w_0=3\) to \(w_0=7\) \gls{rm}. A change in \(\tau_{SP}^-1$
that would compensate for this would involve increasing from the
previous value of $\tau_{SP}^{-1}=0.730\;\text{ns}^{-1}$
($\tau_{SP}=1.37\;\text{ns}$) to
$\tau_{SP}^{-1}=0.878\;\text{ns}^{-1}$
($\tau_{SP}=1.14\;\text{ns}$). It seems reasonable to expect such a
speed-up of rotational diffusion in the less confined environment; the
difference in mobility may correspond to a slight structural difference
in the post-shedding micelle. While further investigations should prove
straightforward, they exceed the scope of this article, which instead
proceeds to analyze different label positions and dispersants.

\subsection{Activation of Interfacial Water Diffusion}

\begin{figure}
\centering
\includegraphics[width=\linewidth]{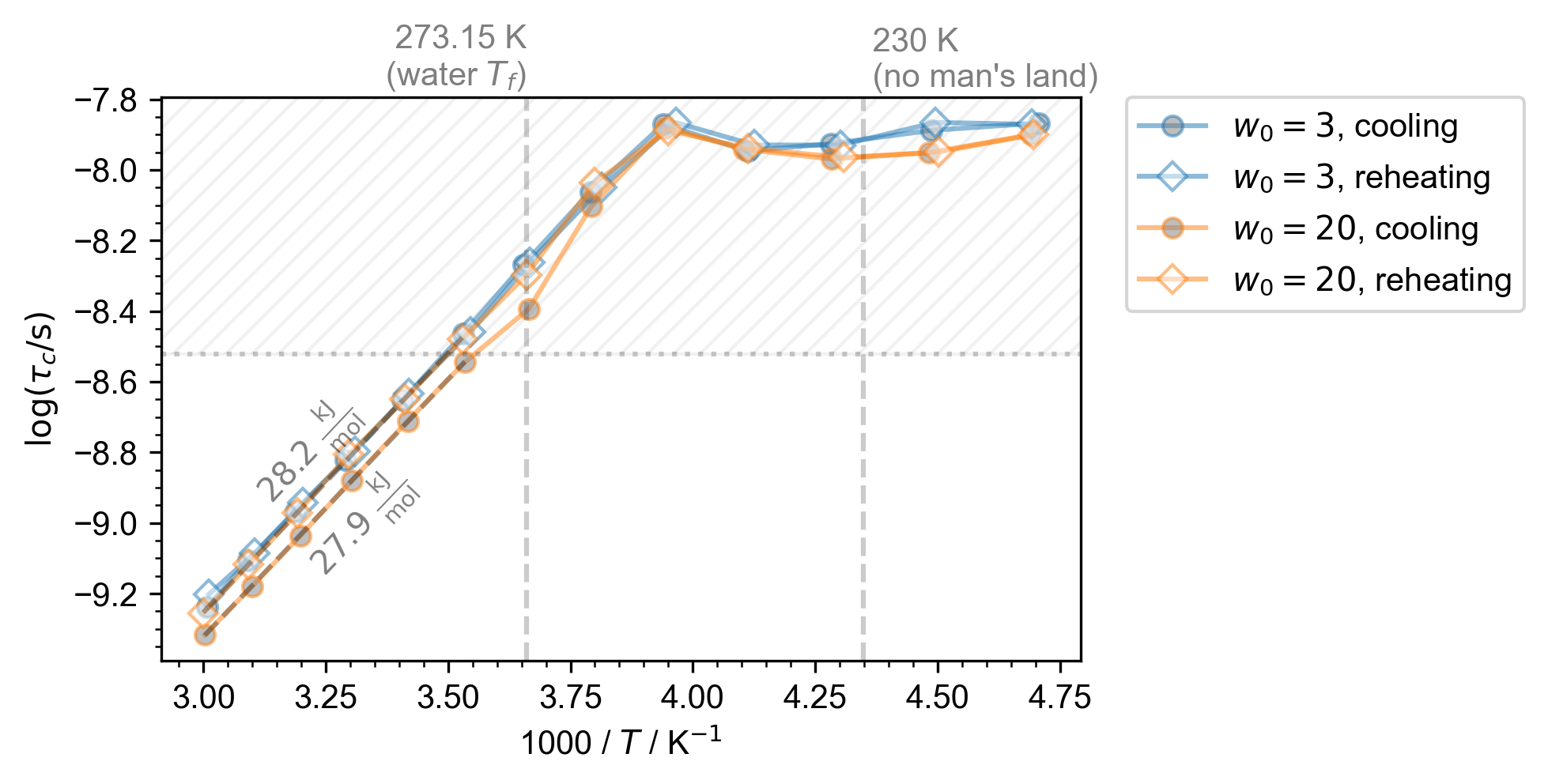}
\caption{Arrhenius temperature dependance of 0.5 mM CAT-16 in isooctane
for ascending and descending temperature
series}\label{fig:ArrhCAT16Isooctane}
\end{figure}

\begin{figure}
\centering
\includegraphics[width=3in]{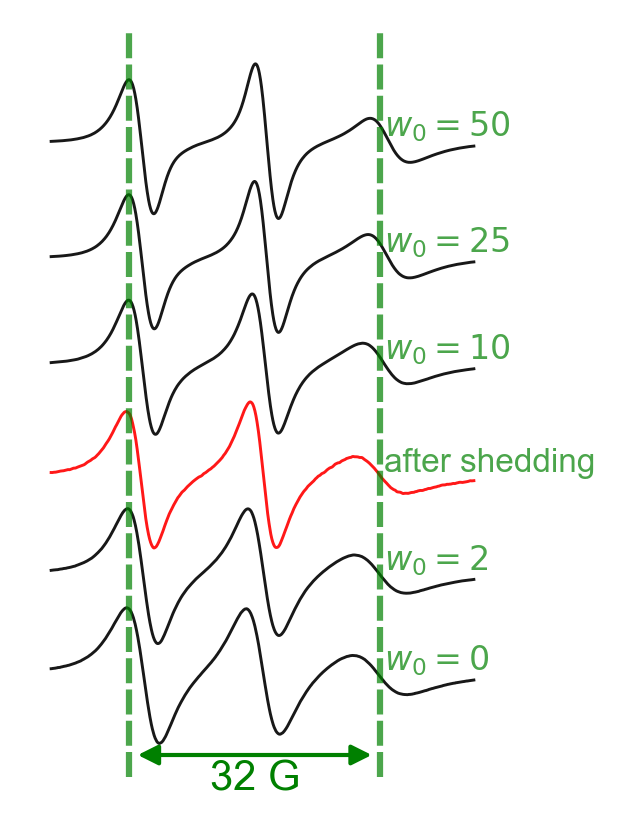}
\caption{\gls{esr} spectra of the spin label CAT-16 at room temperature,
incorporated into \gls{aot} \glspl{rm} with different water loadings ($w_0$). The
red spectrum comes from CAT-16 that was embedded in an \gls{rm} (in
\gls{aot}-isooctane) that was initially $w_0=20$, but which was cycled below
213 K and then reheated to 293 K. As expected for cold-shedding of
water, this lineshape is consistent with smaller
\glspl{rm}.}\label{fig:CAT16_ShedvsW0Series}
\end{figure}

CAT-16 embeds in the interfacial water layer
\cite{Hauser1989IntWatSod} and its rotational diffusion responds to
the diffusivity of water in that layer. Interestingly, at both higher
and lower water loading, the correlation time observed for the
interfacial water is slightly slower than that seen for a temposulfate
embedded in a $w_0 = 3$ \gls{rm} (tbl.~\ref{tbl:Shedvsw0}). A comparison
between the data for the TEMPO-SO\textsubscript{4} and CAT16 spin probes
can also address the question of whether the water contained in the
$w_0=3$ \gls{rm} behaves mostly as interfacial water. The CAT-16 rotation
responds to an activation energy for the diffusivity of water in the
interfacial layer of approximately $28\;\text{kJ}/\text{mol}$, similar
to that observed by the $w_0 = 3$ temposulfate label
($30\;\text{kJ}/\text{mol}$, fig.~\ref{fig:ArrhTEMPOIsooctane}).
Therefore, the $w_0 = 3$ \gls{rm}'s water pool likely consists entirely of
interfacial water molecules. Possibly, the decrease of correlation time
observed with the basic method here (\emph{i.e.},
eq.~\ref{eq:tauFromLinewidth}) upon exchanging spin probes from from
temposulfate to CAT-16 in the $w_0=3$ \gls{rm} could arise from the
molecular-scale order of the CAT-16 relative to the surfactant and the
likely anisotropy of its rotational diffusion tensor.

Finally, the CAT-16 data can aid in estimating the size of the
post-shedding \gls{rm}. Note that for the different water loadings, the
correlation times do vary, consistently with what older literature has
reported \cite{Hauser1989IntWatSod}. This variation as a function of
water loading can help identify the likely water loading of the \gls{rm} that
has experienced shedding. Specifically,
fig.~\ref{fig:CAT16_ShedvsW0Series} compares an (initially $w_0=20$)
\gls{rm} sample that has been subjected to water shedding to a series of \glspl{rm}
with different $w_0$. The sample that has been subjected to shedding
exhibits a lineshape most similar to an untreated sample with a low
water loading of $0<w_0<10$.

\subsection{Freezing in Dodecane Offers Insight into Thermodynamics ofHydration Water}

\begin{figure}
\centering
\includegraphics[width=\linewidth]{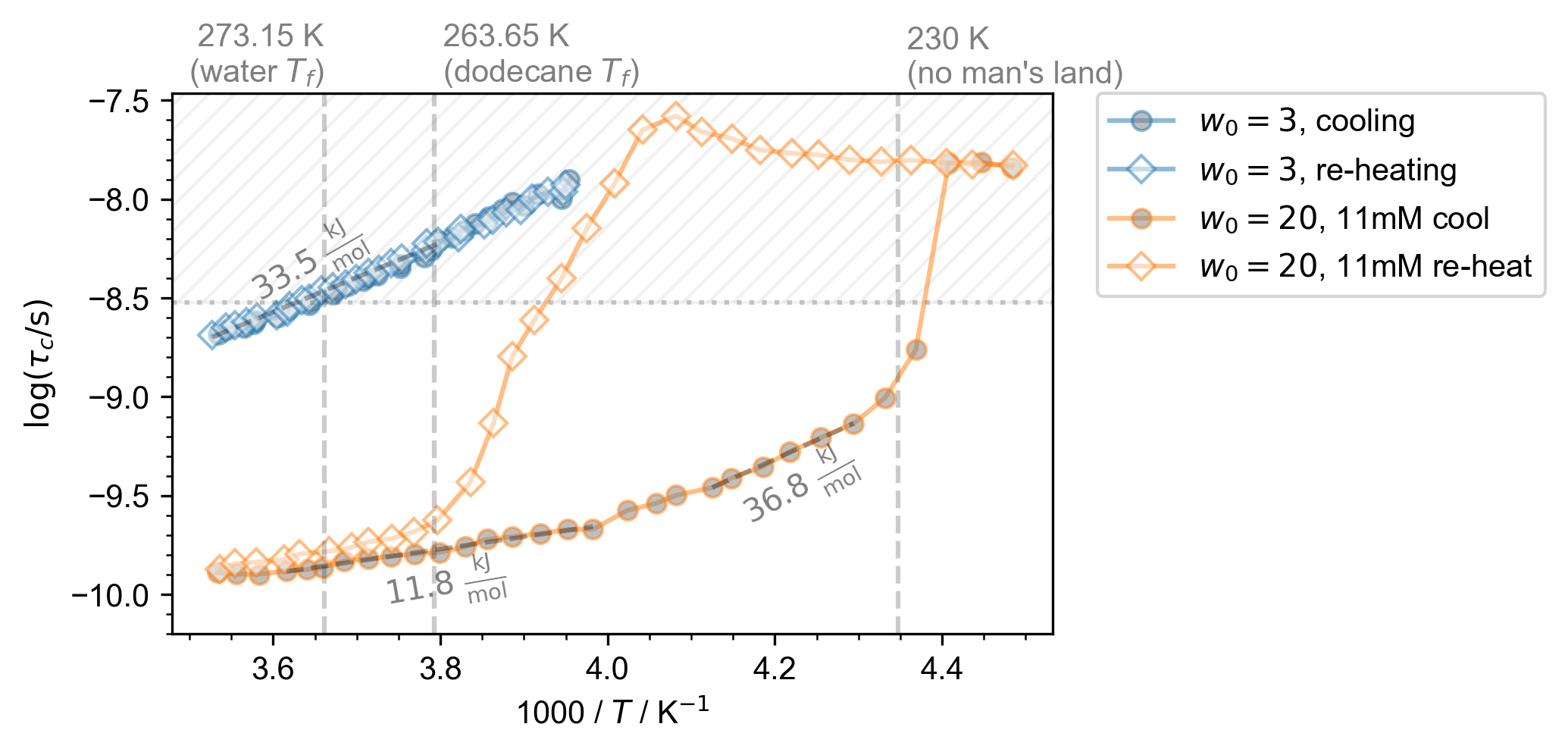}
\caption{Arrhenius temperature dependance of 20 mM
TEMPO-SO\textsubscript{4} in dodecane for ascending or descending
temperature series}\label{fig:ArrhTEMPOdodecane}
\end{figure}

As previously noted, early DSC studies showed freezing of water as low
as 230~K \cite{Boned1986ChaWatDis}, while \gls{nmr} studies showed
freezing of water at temperatures as high as 253~K
\cite{VanHorn2009RevMicEnc,Simorellis2006DynLowTem}; with the latter
result duplicated in the previous section here. The key difference
between these studies appeared to be the dispersant (dodecane \emph{vs.}
isooctane, respectively). Notably, dodecane freezes at 263.65~K, so,
likely, its key property is its ability to immobilize the individual \gls{rm}
aggregates in space and prevent fusion/fission events, without
significantly altering the structure of the \gls{rm} aggregates themselves.
The versatility of cw \gls{esr} enables a controlled study of the dynamics
inside \glspl{rm} dispersed in isooctane (given in previous section) \emph{vs.}
dodecane, which follows.

These measurements of \glspl{rm} in dodecane dispersant focus on a slightly
lower temperature range than in isoctane, in order to enable finer
temperature steps and better temperature equilibration. As for isooctane
(fig.~\ref{fig:ArrhTEMPOIsooctane}) \glspl{rm} with $w_0=3$ dispersed in
dodecane demonstrate a reversible series of spectra with temperature
(fig.~\ref{fig:ArrhCAT16Isooctane}), with a smoothly varying correlation
time. The \gls{aot} binds the water so tightly that the water cannot form ice
crystals, and no freezing transition is observed. This result agrees
with the DSC studies \cite{Boned1986ChaWatDis}, which observed that
water inside \glspl{rm} with $w_0 < 7$ exhibits no freezing transition, even
at temperatures as low as 232~K.

Swapping the isooctane dispersant for dodecane does not significantly
alter the correlation time of the $w_0=3$ sample at 283~K
($10^{-8.70}=2.0\;\text{ns}$ for both dispersants). Since the
viscosities of the two dispersants differ, this result encourages the
conclusion that even for these small \glspl{rm}, and even though the
correlation time is significantly slower than what one expects for a
small spin probe dispersed in bulk water, the rotational diffusion of
the spin probe inside the water encapsulated in the \gls{rm} (\emph{i.e.}
$1/6\tau_{SP}$) likely exceeds (is faster) than the rotational
diffusion of the \gls{rm} as a whole (\emph{i.e.} $1/6\tau_{\gls{rm}}$). The
activation energy of $33.5\;\text{kJ/K}$ is similar to (only slightly
exceeding) the activation energies of either the sulfate probe inside
small \glspl{rm} (\emph{i.e.} $w_0=3$) or the activation energy of the
amphiphilic Cat-16 probe that reports on the interfacial water.

Above $w_0 > 7$, DSC studies \cite{Boned1986ChaWatDis} have
reported a freezing at approximately -41$^{\circ}$C (approximately
232~K), with a heat of freezing roughly proportional to the number of
water molecules above the $w_0=7$ threshold. This temperature
corresponds approximately with the entry into what is commonly denoted
as ``no man's land,'' where ice nucleation proceeds inevitably for all
non-confined environments \cite{Cerveny2016ConWatMod}. Indeed, as
shown in fig.~\ref{fig:ArrhTEMPOdodecane}, the lineshape of
tempo-sulfate changes dramatically as it approaches the boundary to
no-man's land. Specifically, the spin label exhibits much less mobility,
consistent with sharing the nanocomparment with newly formed ice.

\subsection{Freezing in Different Dispersants Offers Insight intoNucleation}

\begin{figure}
\centering
\includegraphics[width=\linewidth]{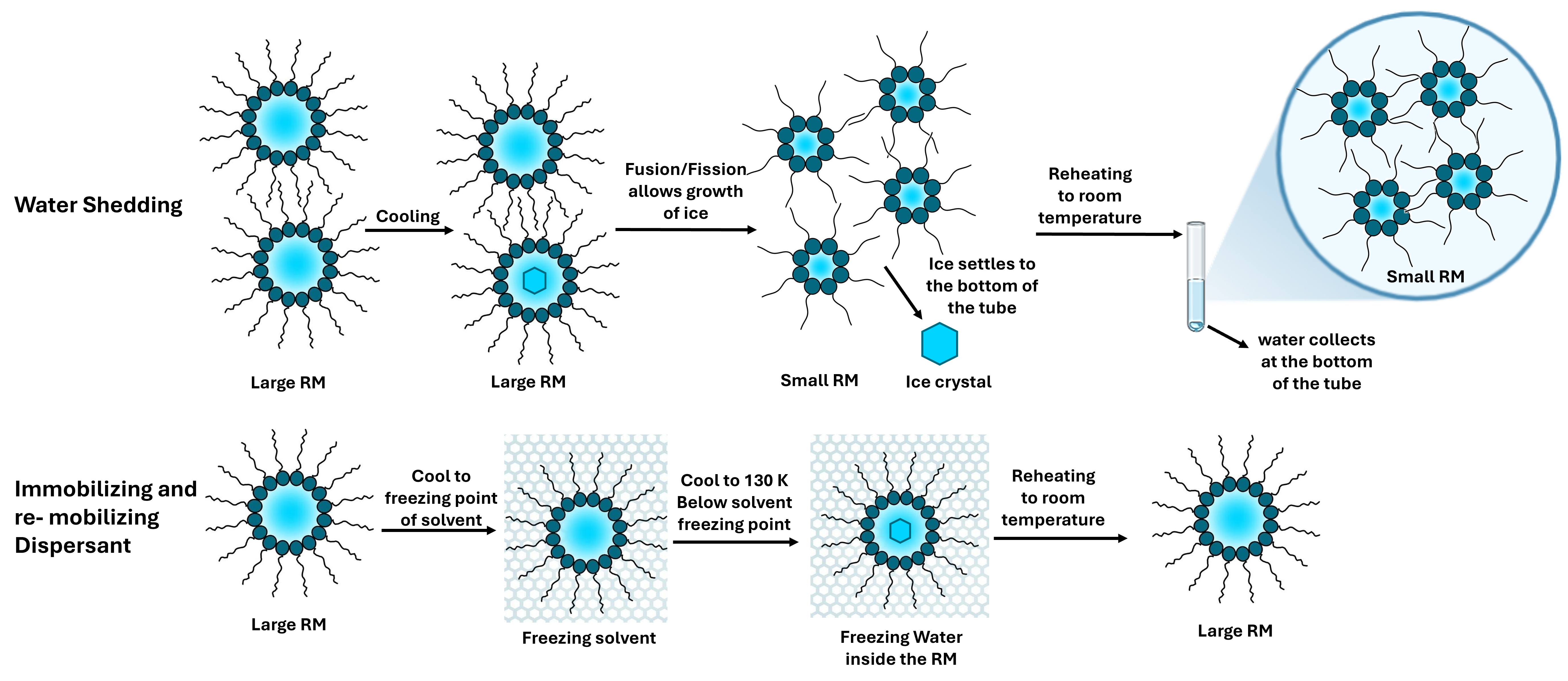}
\caption{Insights into ice nucleation in isooctane and dodecane. (Top)
In isooctane, growth of the ice nucleus proceeds via a fusion-fission
process during cooling. The ice crystals formed during cooling grow past
the critical nucleus size and ultimately leave the \gls{rm}
(``water-shedding''). A smaller \gls{rm} is left behind. Upon re-warming, the
melted ice does not re-enter into the \gls{rm}, but rather collects at the
bottom of the \gls{esr} tube. (Bottom) In dodecane, the solvent freezing
prevents the growth of embryonic ice nuclei inside the \gls{rm}. This causes
the water to remain supercooled until it crosses no-man's land. During
re-warming, the ice melts near 273 K, and the size of the \gls{rm} remains
unchanged.}\label{fig:diff_dispersant}
\end{figure}

The size of the smaller \gls{rm} likely corresponds to the amount of water
that is not ``freezable'' (as opposed to non-crystallizable or
unfreezable bound water). As noted previously
\cite{Simorellis2006DynLowTem} depends on the ionic strength of the
water -- which likely could be phrased as altering the composition of
freezable vs. unfreezable water.

Thus, \glspl{rm} are shown to be a very interesting system for studying
nucleation in confined environments. The \gls{esr} data have already
demonstrated support for the conclusions of older DSC
\cite{Boned1986ChaWatDis} studies: that only the non-interfacial
waters freeze. In dodecane, freezing water molecules manifest as a
hysteresis in the correlation time in (gold lines in
fig.~\ref{fig:ArrhTEMPOIsooctane}). In isooctane, it manifests as a
persistent change to the correlation times arising from the shedding of
the frozen water. The $w_0=3$ samples do not demonstrate either of
these effects, even at temperatures down to no-man's land. In other
words, several water molecules per surfactant are not ``freezable'' --
even at temperatures below no-man's land. Not only can the size of
confinement be changed to explore the influence of the surfactant-water
interface on the ability of the water molecules to freeze, but by
altering the dispersant, the ability of nuclei to grow through
fusion-fission of the \glspl{rm} can also be controlled. Here, isooctane
remains liquid over the temperature range studied, while dodecane
freezes at a temperature slightly lower than bulk water. This difference
significantly affects the behavior of the \glspl{rm} and the freezing dynamics
of the water.

In iso-octane, transient ice nuclei are allowed to grow through
fusion-fission processes between different nanocontainers. Specifically,
a nucleus smaller than the critical radius trapped inside a \gls{rm} with no
remaining ``freezable'' water molecules can grow to the critical size by
taking ``freezable'' water molecules from another \gls{rm} during a fusion
event. Therefore, at temperatures only tens of degrees colder than the
typical freezing point of water, these growing nanocrystals of ice fall
out of the \gls{rm}. Without significant agitation, the water molecules cannot
re-enter the \gls{rm} and are thus irreversibly shed from the \gls{rm}, leaving
behind the surfactant and the spin probe solute.

In dodecane, the properties of the water, in terms of mobility and
polarity appear quite similar. There is no evidence or reasoning that
the water does not still forms the transient nuclei that, in isooctane,
can grow as fusion-fission events bring new non-interfacial water
molecules in contact with the embryonic ice nucleus. However, without
fusion-fission events to supply the new water molecules, the small
nuclei have a high free energy and quickly revert to a liquid state.
This allows water to remain supercooled until it crosses the boundary
into no-man's land at which point it solidifies and requires warming to
near 273~K to melt again.

This controlled comparison between the two dispersants highlights the
conditions under which transient ice nuclei can grow and provides a
deeper understanding of the crystallization process in confined
environments. Roughly, the picture painted by these results appears
consistent with critical nucleation theory
\cite{Lupi2014HetNucIce,Lupi2016PreIntWat,Metya2021IceNucOrg}.
Interestingly, the correlation time measurements, which reflect the
internal viscosity of these nanocontainers, do not appear to respond to
the presence of these transient nuclei, suggesting their rarity, until
possibly $244\;\text{K}$.

\subsection{Limitations of Line Height/Width-Based Methodology}

\begin{figure}
\centering
\includegraphics[width=3in]{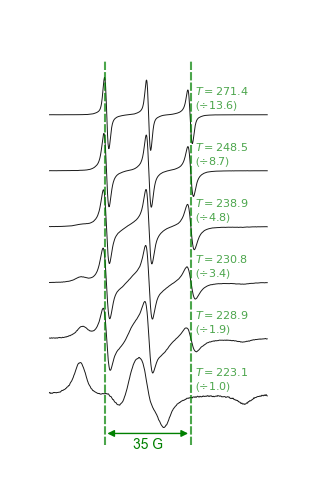}
\caption{Spectra of temposulfate dissolved in supercooled water confined
inside \glspl{rm} suspended in frozen dodecane. As the sample grows colder, an
immobile component appears and grows at the expense of the mobile
spectrum. Because all spectra are normalized the, normalization, the
relative amplitude of each spectrum is given inside
parentheses.}\label{fig:supercooled}
\end{figure}

As previously noted, the rudimentary technique of
eq.~\ref{eq:tauFromLinewidth}, employed in the bulk of this article,
locks onto the three largest peaks in the spectrum. It, of course,
ignores the possibility of multiple components.

Fig.~\ref{fig:supercooled} shows the spectra for \glspl{rm} immobilized in
dodecane, with the signal arising from temposulfate dissolved in
supercooled water (the same as the dark gold circles of
fig.~\ref{fig:ArrhTEMPOdodecane}). Near to 273~K, the spectrum comprises
three near-Lorentzian lines, but, falling temperatures introduce an
immobile component that grows progressively larger. Eventually, at
temperatures approaching the crossover into no man's land, the immobile
component dominates the spectrum.

Recalling the discussion comparing between the liquid isooctane
dispersant (fig.~\ref{fig:ArrhTEMPOIsooctane}) and the frozen dodecane
dispersant (fig.~\ref{fig:ArrhTEMPOdodecane}), this observation is
tantalizing. Specifically, if transient nuclei were being formed inside
the immobilized \glspl{rm}, one would expect the spectra of temposulfate to
reflect this to a certain degree. And this is exactly what
fig.~\ref{fig:supercooled} provides. As the temperature grows colder and
colder, the likelihood of ice nuclei formation, as well as the formation
of larger-sized ice nuclei, becomes greater and greater. Consequently,
the immobile component forms a greater proportion of the overall \gls{esr}
spectrum.

\section{Conclusions}

Another particularly interesting property of certain \gls{rm} preparations
(notably those in dodecane and cyclohexane) is that they can freeze,
locking down the 3D translational motion of the \gls{rm} aggregates. This
offers the unique opportunity to create nano-compartments that remain
non-interacting and spatially fixed, yet, that can be released by
activating the phase transition. Here, \gls{esr} offers insight into whether
or not the water and/or surfactant interface of the \gls{rm} aggregates
maintains some amount of rotational mobility.

While the study of supercooled water is both deep and has great breadth,
it is interesting to note that this particular set of observations
preliminarily could be explained with a very simplistic picture
involving the transient formation of ice nuclei inside the confined
environment.

The hydrogen-bonded matrix is what interacts with macromolecular systems
like polymers, nanoparticles, or proteins. Macromolecules do not
interact with individual water molecules but with the entire
hydrogen-bonded network. The key takeaway is that the hydrogen-bonded
matrix is more than the sum of its parts. While this concept isn't new,
this is a specific example that highlights the crucial role of the
hydrogen-bonded matrix in macromolecular chemistry.

\appendix
\ifarxiv
\putbib
\end{bibunit}
\pagebreak
\onecolumngrid
\begin{center}
\makeatletter
\textbf{\large Supplemental Materials for:\\
\@title}
\makeatother
\end{center}
\twocolumngrid
\ifhassi
\begin{bibunit}
\setcounter{equation}{0}
\setcounter{figure}{0}
\setcounter{table}{0}
\setcounter{page}{1}
\renewcommand{\theequation}{S\arabic{equation}}
\renewcommand{\thefigure}{S\arabic{figure}}
\renewcommand\thesubfigure{S\arabic{tempfigure}\alph{subfigure}}
\renewcommand{\thesection}{S\arabic{section}}
\renewcommand{\thepage}{S\arabic{page}}
\renewcommand{\bibnumfmt}[1]{[S#1]}
\renewcommand{\citenumfont}[1]{S#1}
\glsresetall
\newcommand{\maincref}[1]{\cref{#1} (main text)}
\section{Supporting Information}

\begin{figure}
\hypertarget{fig:HaeringCompositeFig}{%
\centering
\includegraphics[width=3.5in]{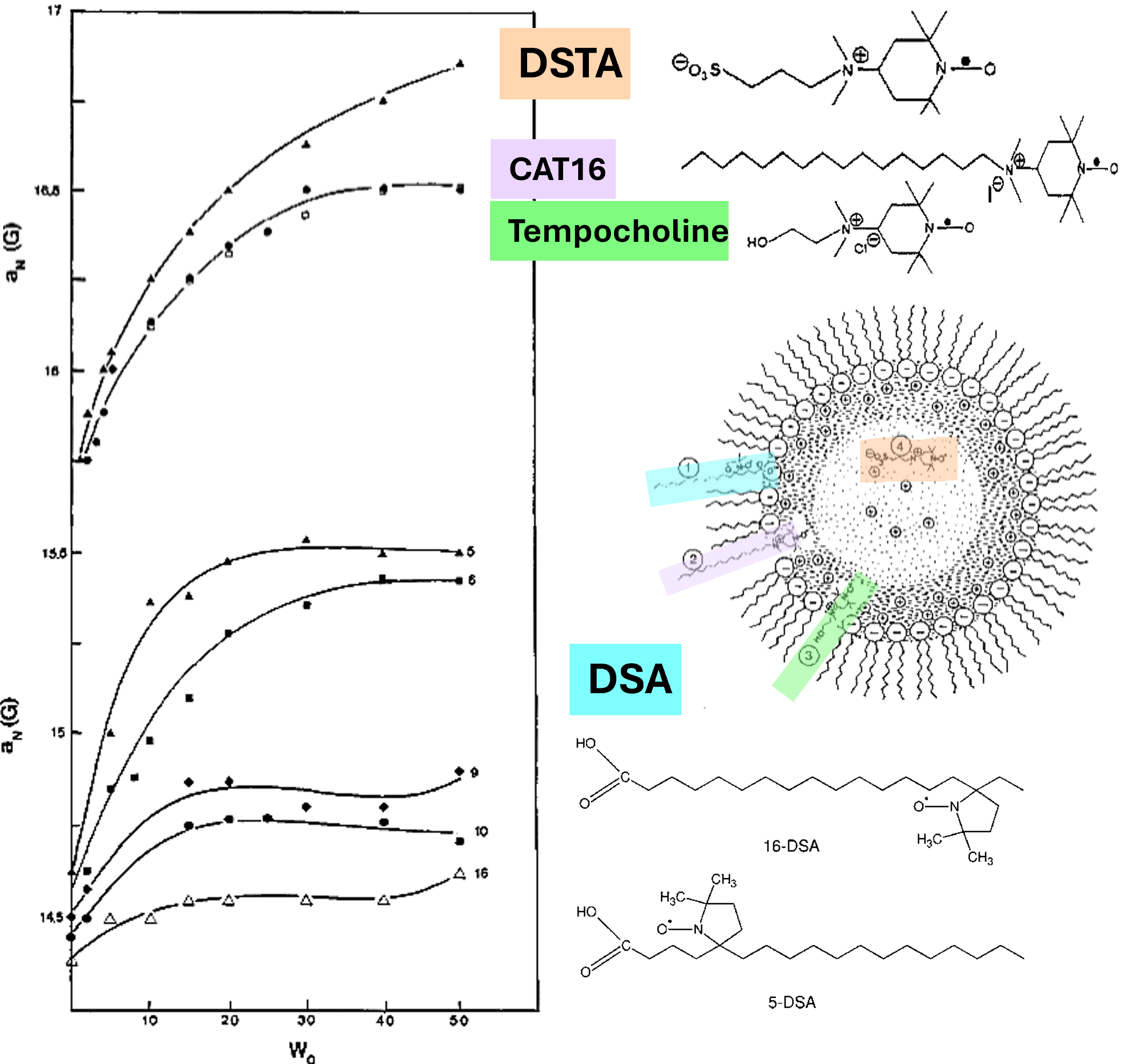}
\caption{Isotropic hyperfine splitting constant ($a_N$) in gauss as a
function of water loading ($W_0$) for different spin labels in the \gls{aot}
\gls{rm}, adapted from
Haering.\cite{Haering1988ChaEleSpi}}\label{fig:HaeringCompositeFig}
}
\end{figure}

\hypertarget{tbl:Haering}{}
\begin{longtable}[]{@{}
  >{\centering\arraybackslash}p{(\columnwidth - 8\tabcolsep) * \real{0.1000}}
  >{\centering\arraybackslash}p{(\columnwidth - 8\tabcolsep) * \real{0.2222}}
  >{\centering\arraybackslash}p{(\columnwidth - 8\tabcolsep) * \real{0.0889}}
  >{\centering\arraybackslash}p{(\columnwidth - 8\tabcolsep) * \real{0.2333}}
  >{\raggedright\arraybackslash}p{(\columnwidth - 8\tabcolsep) * \real{0.3222}}@{}}
\caption{\label{tbl:Haering}correlation time ($\tau$) of aqueous spin
label DSTA and surfactant-water interface spin label
CAT-16.}\tabularnewline
\toprule\noalign{}
\multicolumn{2}{@{}>{\centering\arraybackslash}p{(\columnwidth - 8\tabcolsep) * \real{0.3222} + 2\tabcolsep}}{%
\begin{minipage}[b]{\linewidth}\centering
DSTA
\end{minipage}} &
\multicolumn{2}{>{\centering\arraybackslash}p{(\columnwidth - 8\tabcolsep) * \real{0.3222} + 2\tabcolsep}}{%
\begin{minipage}[b]{\linewidth}\centering
CAT-16
\end{minipage}} & \begin{minipage}[b]{\linewidth}\raggedright
Ratio of correlation times
\end{minipage} \\
\begin{minipage}[b]{\linewidth}\centering
$w_0$
\end{minipage} & \begin{minipage}[b]{\linewidth}\centering
$\tau_{DSTA}$ / ns
\end{minipage} & \begin{minipage}[b]{\linewidth}\centering
$w_0$
\end{minipage} & \begin{minipage}[b]{\linewidth}\centering
$\tau_{Cat16}$ / ns
\end{minipage} & \begin{minipage}[b]{\linewidth}\raggedright
$\tau_{DSTA}/\tau_{Cat16}$
\end{minipage} \\
\midrule\noalign{}
\endfirsthead
\toprule\noalign{}
\multicolumn{2}{@{}>{\centering\arraybackslash}p{(\columnwidth - 8\tabcolsep) * \real{0.3222} + 2\tabcolsep}}{%
\begin{minipage}[b]{\linewidth}\centering
DSTA
\end{minipage}} &
\multicolumn{2}{>{\centering\arraybackslash}p{(\columnwidth - 8\tabcolsep) * \real{0.3222} + 2\tabcolsep}}{%
\begin{minipage}[b]{\linewidth}\centering
CAT-16
\end{minipage}} & \begin{minipage}[b]{\linewidth}\raggedright
Ratio of correlation times
\end{minipage} \\
\begin{minipage}[b]{\linewidth}\centering
$w_0$
\end{minipage} & \begin{minipage}[b]{\linewidth}\centering
$\tau_{DSTA}$ / ns
\end{minipage} & \begin{minipage}[b]{\linewidth}\centering
$w_0$
\end{minipage} & \begin{minipage}[b]{\linewidth}\centering
$\tau_{Cat16}$ / ns
\end{minipage} & \begin{minipage}[b]{\linewidth}\raggedright
$\tau_{DSTA}/\tau_{Cat16}$
\end{minipage} \\
\midrule\noalign{}
\endhead
\bottomrule\noalign{}
\endlastfoot
0 & 5.89 & 0 & 5.29 & 1.11 \\
2 & 2.44 & 2 & 1.55 & 1.57 \\
5 & 1.59 & & & \\
10 & 0.899 & & & \\
50 & 0.308 & 50 & 1.4 & 0.22 \\
\end{longtable}

\begin{figure}
\hypertarget{fig:CaldararuTau}{%
\centering
\includegraphics[width=\linewidth]{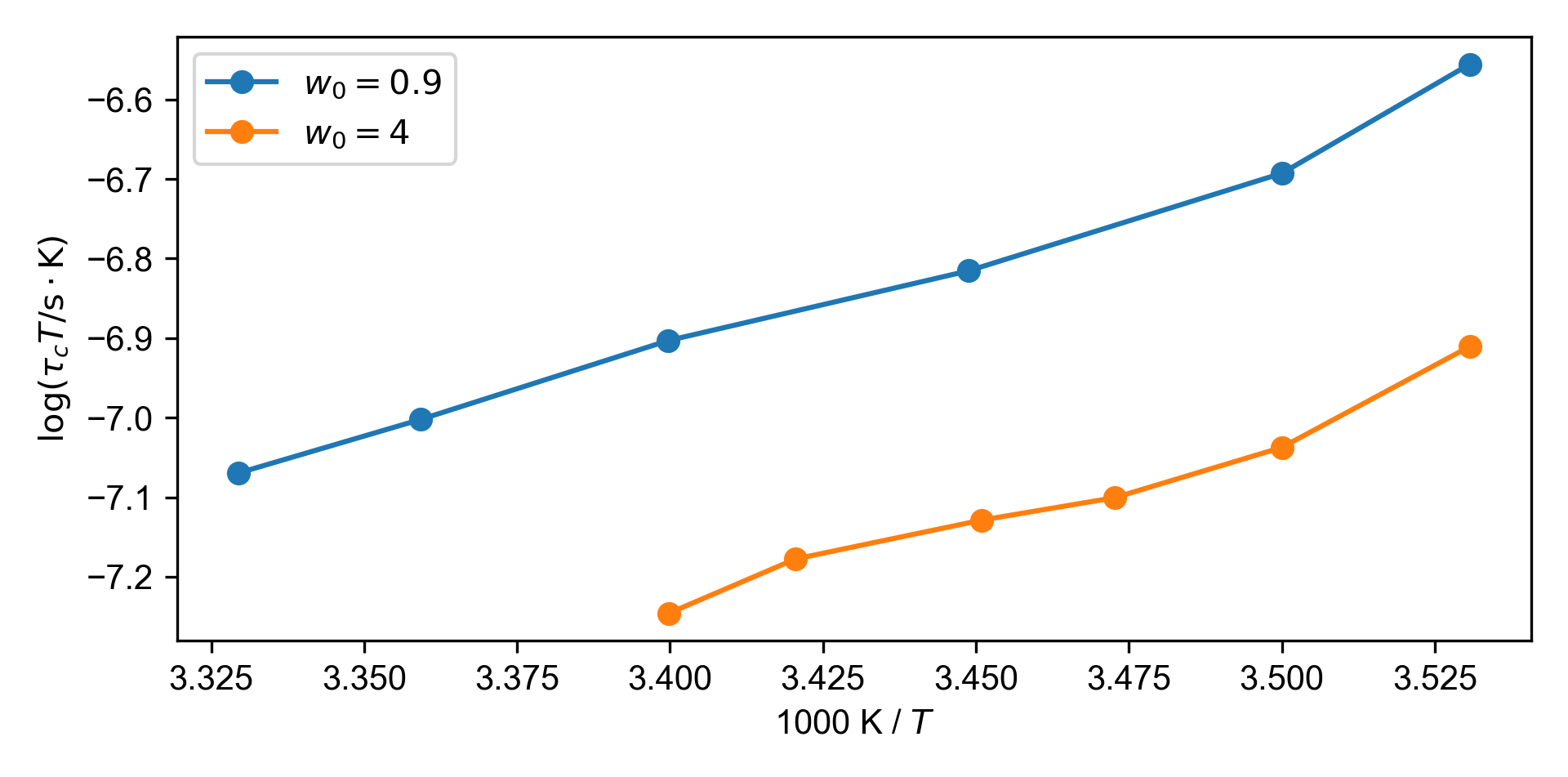}
\caption{Arrhenius temperature dependance for CAT 4 in
C\textsubscript{12}E\textsubscript{4}/cyclohexane/water system for low
water loading ($w_0$). (Digitized and converted to arrhenius
temperature dependance from the data presented in the figures by
Caldararu\cite{Caldararu1998StrAspSel})}\label{fig:CaldararuTau}
}
\end{figure}

\subsubsection{Characterization of Synthesized spin labels}

\begin{figure}
\hypertarget{fig:NMR_TEMPOSO4}{%
\centering
\includegraphics[width=\linewidth]{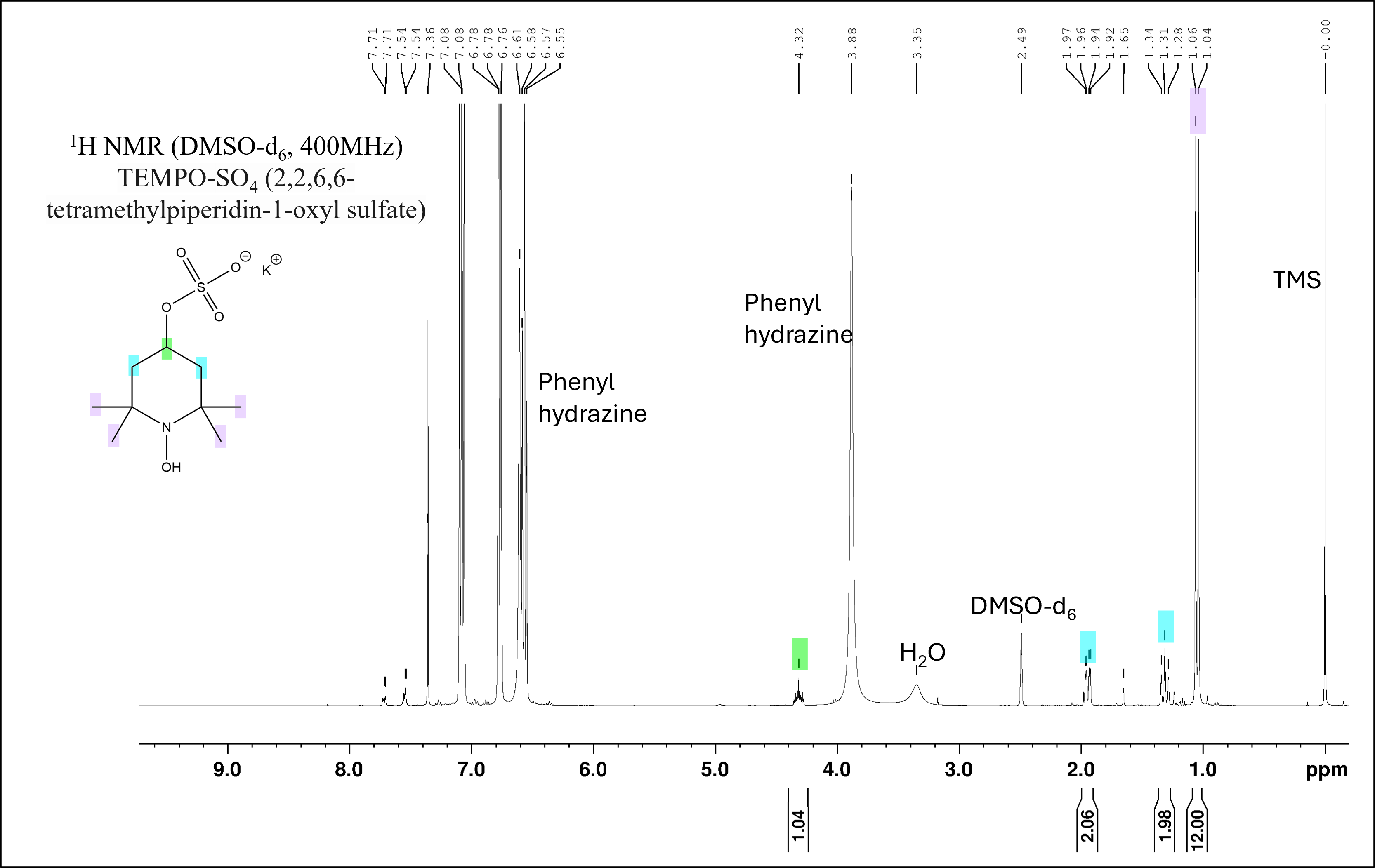}
\caption{\gls{nmr} of TEMPO-SO\textsubscript{4}}\label{fig:NMR_TEMPOSO4}
}
\end{figure}

\begin{figure}
\hypertarget{fig:NMR_CAT16}{%
\centering
\includegraphics[width=\linewidth]{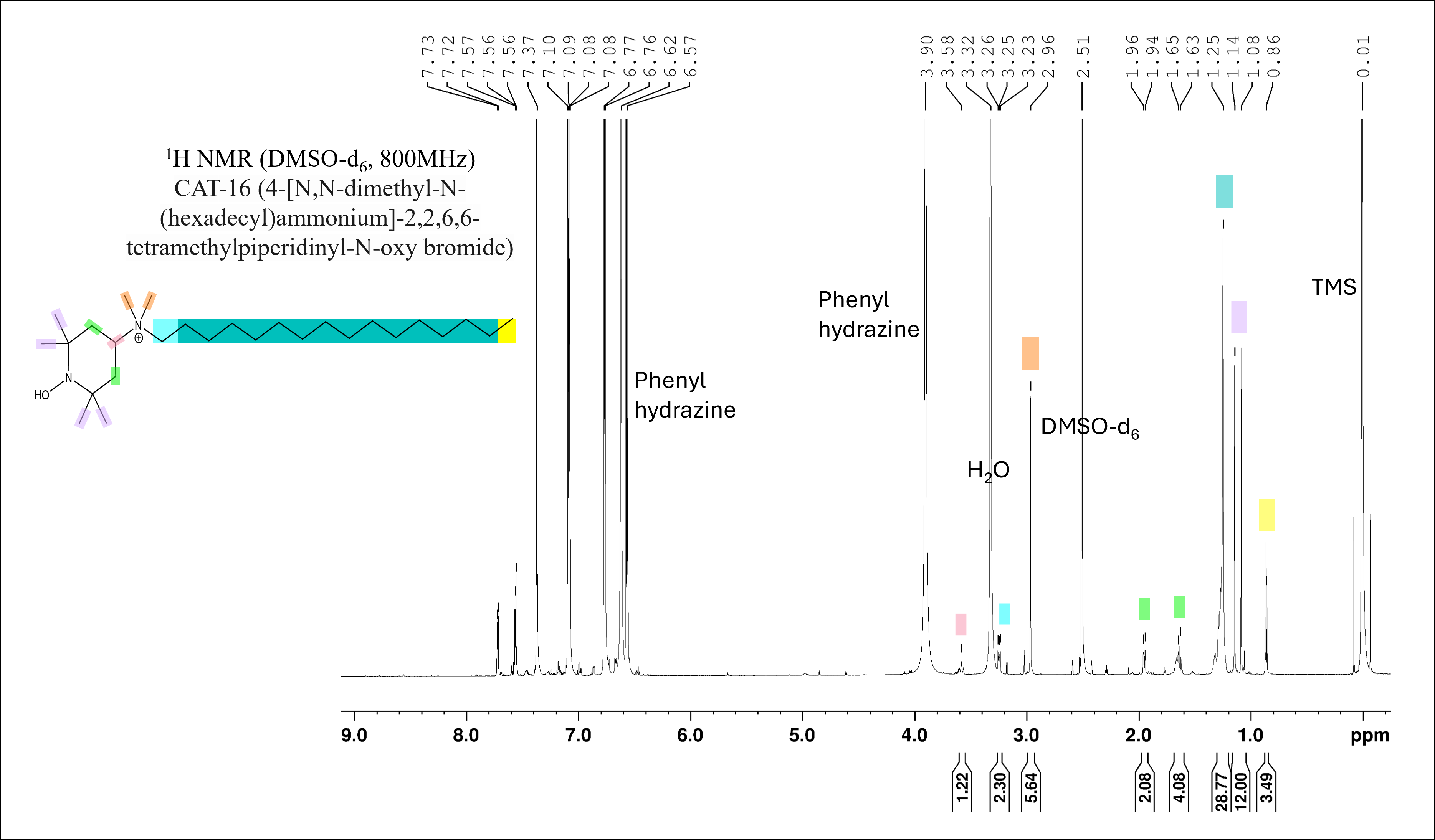}
\caption{\gls{nmr} of CAT-16}\label{fig:NMR_CAT16}
}
\end{figure}

fig.~\ref{fig:NMR_TEMPOSO4} and fig.~\ref{fig:NMR_CAT16} give the \gls{nmr}
spectra for both the phenylhydrazine-quenched spin labels used in this
study, with a color-coding to indicate the assignment.

\subsection{Concisely Illustrating Spectral Variation over a Series}

Throughout this paper, when selecting spectra out of a temperature
series for presentation, a strategy for optimally choosing 6-8
temperatures was pursued. Specifically the function: \[f(T) = \frac{
    \sqrt{\int
        \left|\frac{\partial s(B_0, T)}{\partial T}\right|^2 dB_0}
    }{
    \sqrt{\int \left| s(B_0, T) \right|^2 \, dB_0}
}\] was developed to quantify how the shape of the spectrum
($s(B_0,T)\)) varies with temperature (\(T$). By identifying the
temperatures that yield equally spaced y-values on the plot of a
cumulative integral of $f(T)$, one can choose a set of representative
spectra whose shape differs roughly equally from each other. For a such
a strategic set of spectra at different temperatures, starting from the
(previously described) lowest temperature, the best fit parameters for
each fit, when employed as starting values for the fit at the next
highest temperature, successfully yield high-fidelity fits for the next
highest temperature in the series.

\subsection{Correlation time in post-shedding micelle.}

To get a sense of the size of the post-shedding \gls{rm}, we can compare the
correlation time of the Cat-16 label to that in other sized \glspl{rm}.

\hypertarget{tbl:Shedvsw0}{}
\begin{longtable}[]{@{}
  >{\centering\arraybackslash}p{(\columnwidth - 2\tabcolsep) * \real{0.2778}}
  >{\raggedright\arraybackslash}p{(\columnwidth - 2\tabcolsep) * \real{0.2917}}@{}}
\caption{\label{tbl:Shedvsw0}Correlation time ($\tau_c$) of the spin
probe CAT-16 in different water loadings, as shown in
fig.~\ref{fig:CAT16_ShedvsW0Series}. After the water-shedding, the red
spectrum in fig.~\ref{fig:CAT16_ShedvsW0Series} shows a slower
correlation time (referred to as after shedding)}\tabularnewline
\toprule\noalign{}
\multicolumn{2}{@{}>{\centering\arraybackslash}p{(\columnwidth - 2\tabcolsep) * \real{0.5694} + 2\tabcolsep}@{}}{%
\begin{minipage}[b]{\linewidth}\centering
CAT-16
\end{minipage}} \\
\begin{minipage}[b]{\linewidth}\centering
$w_0$
\end{minipage} & \begin{minipage}[b]{\linewidth}\raggedright
$\tau_c$ / ns
\end{minipage} \\
\midrule\noalign{}
\endfirsthead
\toprule\noalign{}
\multicolumn{2}{@{}>{\centering\arraybackslash}p{(\columnwidth - 2\tabcolsep) * \real{0.5694} + 2\tabcolsep}@{}}{%
\begin{minipage}[b]{\linewidth}\centering
CAT-16
\end{minipage}} \\
\begin{minipage}[b]{\linewidth}\centering
$w_0$
\end{minipage} & \begin{minipage}[b]{\linewidth}\raggedright
$\tau_c$ / ns
\end{minipage} \\
\midrule\noalign{}
\endhead
\bottomrule\noalign{}
\endlastfoot
0 & 2.13 \\
2 & 2.05 \\
10 & 1.92 \\
25 & 1.87 \\
50 & 1.87 \\
after shedding & 2.22 \\
\end{longtable}

\putbib
\end{bibunit}
\fi
\else
\bibliography{references}
\fi
\end{document}